\documentstyle[11pt,fullpage,doublespace,epsf]{article}

 \DeclareMathSizes{11}{19}{13}{9}   

\makeatletter
\@addtoreset{equation}{section}
\makeatother

\begin{document}

   \title{ QCD Cosmology from the Lattice Equation of State}

\author{Michael McGuigan and Wolfgang S\"oldner\\Brookhaven National Laboratory\\Upton NY 11973\\mcguigan@bnl.gov}
\date{}
\maketitle

\begin{abstract}
We numerically determine the time dependence of the scale factor from the lattice QCD equation of state, which can be used to define a QCD driven cosmology.  We compare a lattice approach to QCD cosmology at late times with other models of  the low temperature equation of state including the hadronic resonance gas model, Hagedorn model and  AdS/CFT.
\end{abstract}

\section{Introduction}

QCD cosmology is important as the Universe exits the radiation dominated era and enters the matter dominated era, eventually evolving into the dark energy dominated era \cite{Witten:1984rs,Schwarz:2003du,Muller,Borghini:2000yp,Kapusta:2000fe,Chandra:1999tr,Coley:1993zg,Suganuma:1996yb,Olive:1990rd}. The effect of the strong interactions on cosmology was considered early on \cite{Hagedorn:1965st,Huang:1970iq} but the nonperturbative nature of the strong interactions at low energy limited the progress of the subject. Today with nonperturbative approaches like lattice QCD \cite{Cheng:2007jq} and AdS/QCD \cite{Gubser:2008ny,Gubser:2008yx,Gubser:2008sz} it is important to revisit the subject to see what can be said quantitatively about QCD cosmology.
Also besides the time dependence of the scale factor in QCD cosmology, one can also look at the effect of recent advances on the QCD equation of state for exotic astrophysical objects like quark or strange stars \cite{Freedman:1977gz,Chakrabarty:1991ui}.

In this paper we consider a spatially flat Friedmann-Robertson-Walker Universe where the radius of the Universe $a(t)$, energy density $\varepsilon$, and pressure $p$ all depend on time. Conservation of the local energy-momentum tensor leads to:
\[
\frac{{d(\varepsilon a^3 )}}{{dt}} =  - p\frac{{d(a^3 )}}{{dt}}.
\label{eq:emtensor}
\]
The Friedmann equation relates the time evolution of the Universe to energy density and is is given by:
\begin{equation}
3M_P^2 (\frac{1}{a}\frac{{da}}{{dt}})^2  = \varepsilon
\end{equation}
where $M_P = \sqrt{1/8\pi G_N}$ is the reduced Planck mass. The equation of state refers to the dependence of energy density and pressure on temperature $T$. For a given equation of state one can rewrite the above two equations in a form which determines the expansion factor $a(t)$.
First we can determine the time dependence of the temperature  from:
\begin{equation}
\frac{{dT}}{{dt}} =  - 3(\varepsilon (T) + p(T))\sqrt {\frac{{ \varepsilon (T)}}{3M_P^2}} (\frac{{d\varepsilon (T)}}{{dT}})^{ - 1} .
\label{eq:dTdt}
\end{equation}
This can be integrated to give:
\begin{equation}
t(T) = \int_T^{T_{\max } } {d\bar T} \frac{{\frac{{d\varepsilon (\bar T)}}{{d\bar T}}}}{{3(\varepsilon (\bar T) + p(\bar T))\sqrt {\frac{{\varepsilon (\bar T)}}{3M_P^2}} }}.
\label{eq:tT}
\end{equation}
Then one can invert the function $t(T)$ to yield $T(t)$. Next one can
determine the radius $a(t)$ time dependence from
equation~(\ref{eq:emtensor}) which is integrated to yield:
\begin{equation}
  a(t) = \exp (\frac {1}{\sqrt{3}M_P} \int_{t_0 }^t dt {\sqrt {\varepsilon (T(t))} } ).
\label{eq:at}
\end{equation}

For a simple example consider a radiation dominated universe with equation of state:
\[
\begin{array}{l}
 \varepsilon (T) = \alpha_{rad} T^4,  \\ 
 p(T) = \frac{\alpha_{rad} }{3}T^4.
 \end{array}
\]
where $\alpha_{rad}$ is a constant. Solving equation~(\ref{eq:dTdt}) we
find $T(t) \propto t^{-1/2}$ and $\varepsilon (T(t)) = 3M_P^2
\frac{1}{{4t^2 }}$.  Performing the integration (\ref{eq:at}) we obtain:
\[
a(t) = \exp (\frac{1}{2}\log (t)) = t^{1/2}.
\]
This $t^{1/2}$ time dependence also  gives the very high temperature and early time behavior of QCD cosmology.

For nonzero cosmological constant $\lambda$ the Friedmann equation is modified to:
\[
3M_P^2(\frac{1}{a}\frac{{da}}{{dt}})^2  = \varepsilon (T(t))+\lambda.
\]
For all the models of low energy QCD that we use in this paper the
cosmological constant provides only a small modification. This is
because the energy density scale of low energy QCD is typically $(100
MeV)^4$ whereas the cosmological constant is $(2.3
meV)^4$~\cite{Yao:2006px}.  Modifications are
expected by the cosmological constant as can been seen, e.g., by the
scale factor for matter domination
\[
a(t) = (\sinh (\frac{{\sqrt{3}\sqrt \lambda  }}{{2 M_P }}t))^{2/3} 
\]
which reduces to $t^{2/3}$ at small times and $\exp
(\frac{\sqrt{\lambda}}{\sqrt{3} M_P}t )$ at large times. We will not
investigate corrections with respect to the cosmological constant in
this paper and leave this for further study at a later stage.

Another possible modification to the cosmological equations for $a(t)$ involves the bulk viscosity. The inclusion of bulk viscosity modifies the conservation equation by:
\[
\frac{{d(\varepsilon a^3 )}}{{dt}} =  - p\frac{{d(a^3 )}}{{dt}} + 
\zeta \frac{1}{a^3}(\frac{{d( a^3 )}}{{dt}})^2 
\]
where $\zeta$ is the bulk viscosity. QCD bulk viscosity is currently
being studied in heavy ion collisions and QCD, see
e.g.~\cite{Luzum:2008cw} and~\cite{Kharzeev:2007wb}, respectively,
thus it is of interest to see what cosmological effect the bulk viscosity may have.
We leave the study of the cosmological effects of the bulk viscosity
for future work.

For any realistic equation of state related to QCD the above steps which determine $t(T)$, $T(t)$, $\varepsilon(T(t))$ and $a(t)$ all have to be done numerically. We shall find a simple radiation dominated  picture for the scale factor $a(t)$ at early times corresponding to deconfinement. At late times the time dependence of the scale factor is quite complex. One can use an effective description using masses of various resonances form the particle data table as in the Hadronic Resonance Gas Model (HRG)\cite{Karsch:2003vd,Karsch:2003zq,SakthiMurugesan:1990pe,Tawfik:2004sw}, one can use an effective Hagedorn string picture \cite{Hagedorn:1965st} as Weinberg and Huang did in \cite{Huang:1970iq}, one can use a AdS/CFT approach to the equation of state \cite{Gubser:2008ny}, a symmetry which relates small radius to large radius QCD-like theories if one thinks of the time direction as compactified as in an imaginary time formalism \cite{Shifman:2008ja}, or one can use lattice simulations to determine the equation of state \cite{Cheng:2007jq}. In this paper we compare the results of some of these approaches applied to the calculation of the scale factor $a(t)$.

This paper is organized as follows. In section 2 we discuss the numerical determination of the scale factor of the Universe as a function of time, $a(t)$, from the lattice QCD equation of state. We work with zero chemical potential and bulk viscosity. We plan to include these in future work. In section 3 we determine the scale factor from the hadronic resonance gas model  model.  In section 4 we discuss the scale factor time dependence for an early Universe model which has a Hagedorn ultimate temperature. In section 5 we discuss the scale factor determined from the AdS/CFT approach to the equation of state following a similar approach to Gubser and Nellore \cite{Gubser:2008ny}. In section 6 we state the main conclusions of the paper.

\section{Lattice QCD equation of state}

\begin{figure}[tp]

   \centerline{\hbox{
   \epsfxsize=3in
   \epsffile{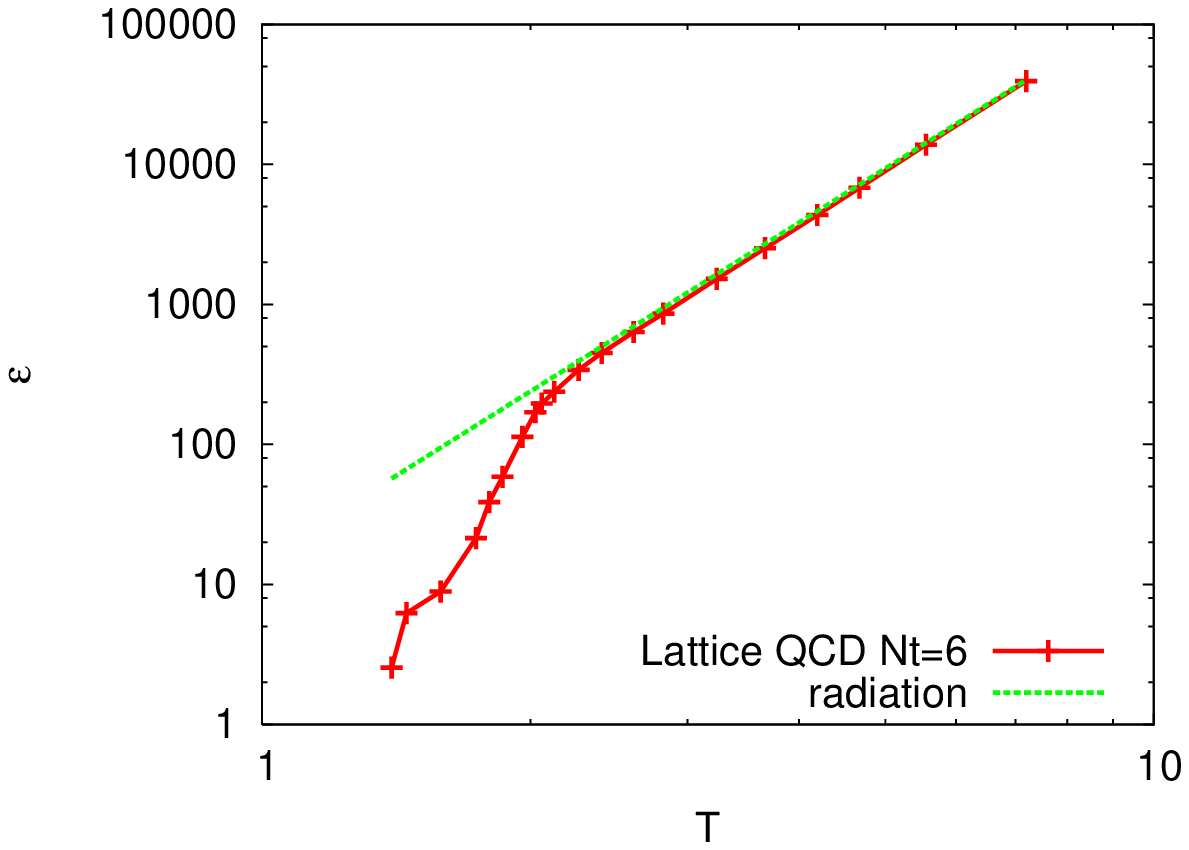}
   \epsfxsize=3in
   \epsffile{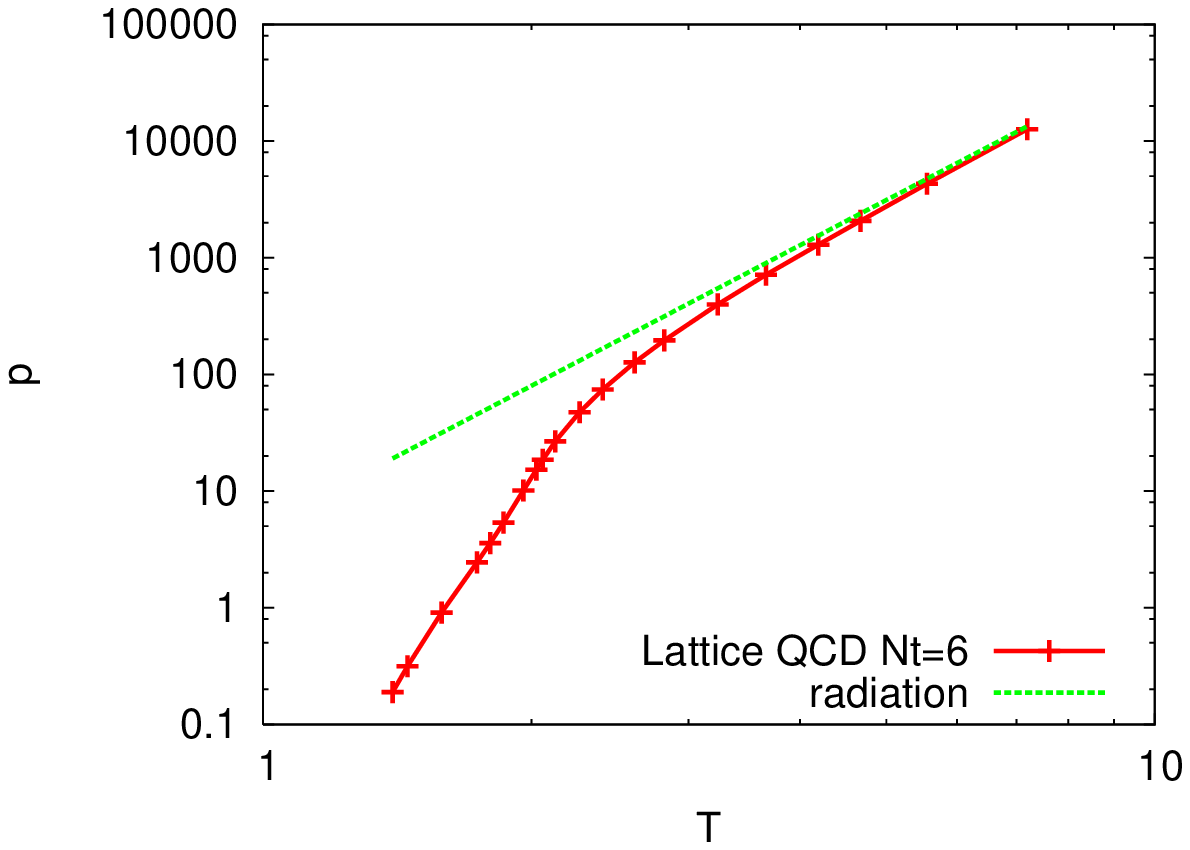}
     }
  }
  \caption{Plot of energy density (left) and pressure (right) versus
    temperature from Lattice QCD data with $N_t=6$. Continuous line is
    an interpolation through the data. For comparison, we have plotted
    the expected behavior for radiation.}

  \label{fig1}

\end{figure}

Lattice QCD is a modern tool which allows one to systematically study
the non-perturbative regime of the QCD equation of state.  Utilizing
supercomputers the QCD equation of state was calculated on the lattice
in~\cite{Cheng:2007jq} with two light quarks and a heavier strange
quark on a $(N_t=6)\times 32^3$ size lattice. The quark masses have been chosen to be close to their physical
value, i.e. the pion mass is about $220 MeV$. For further details we
refer the reader to~\cite{Cheng:2007jq}, however, we like to remark
that the equation of state was calculated at a temporal extent of the
lattice $N_t=6$ and for $N_t=6$ sizable lattice cut-off effects are
still present~\cite{Gupta:2008}. 
\begin{figure}[tbp]
   \centerline{\hbox{
   \epsfxsize=3.0in
   \epsffile{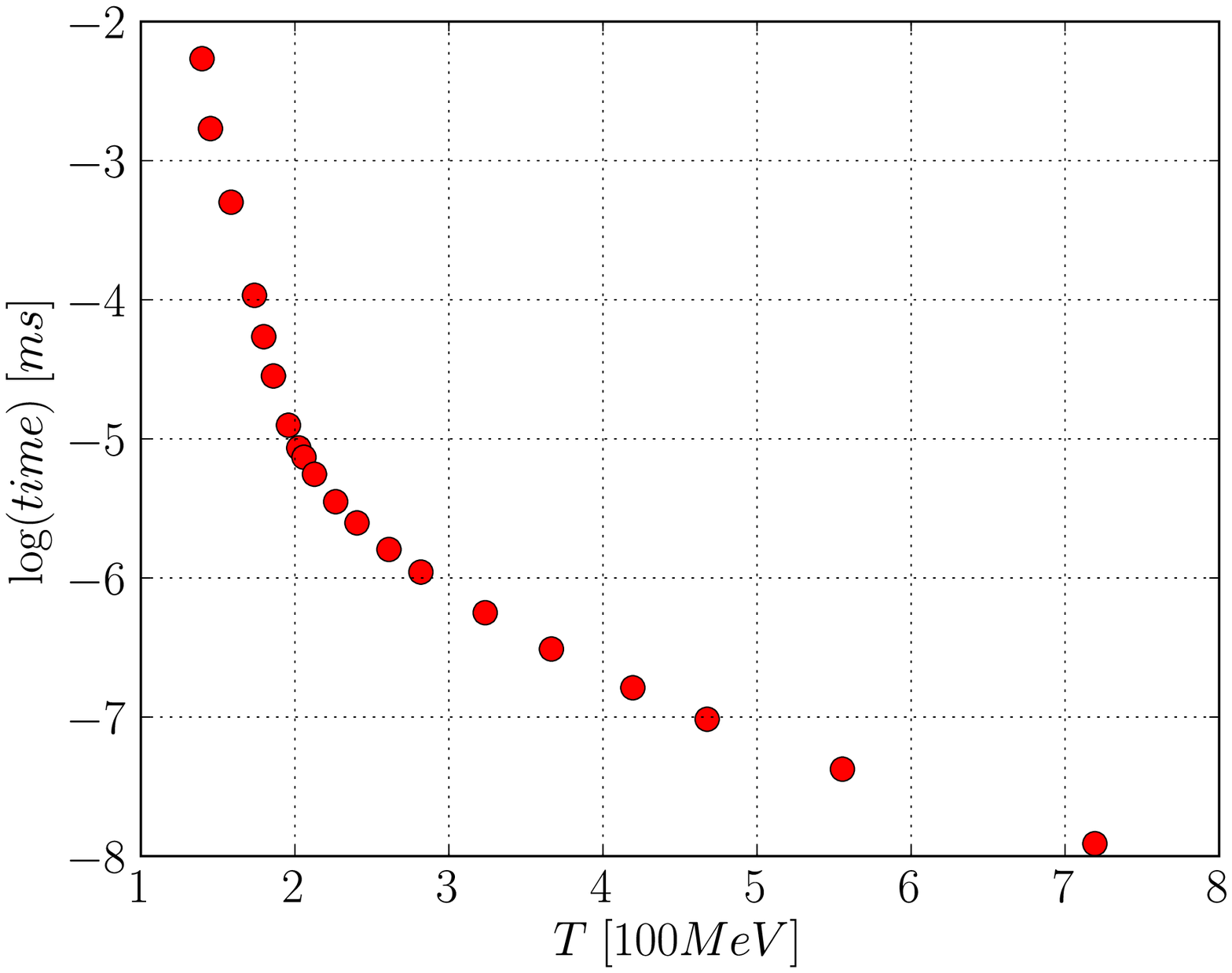}
   \epsfxsize=3.0in
   \epsffile{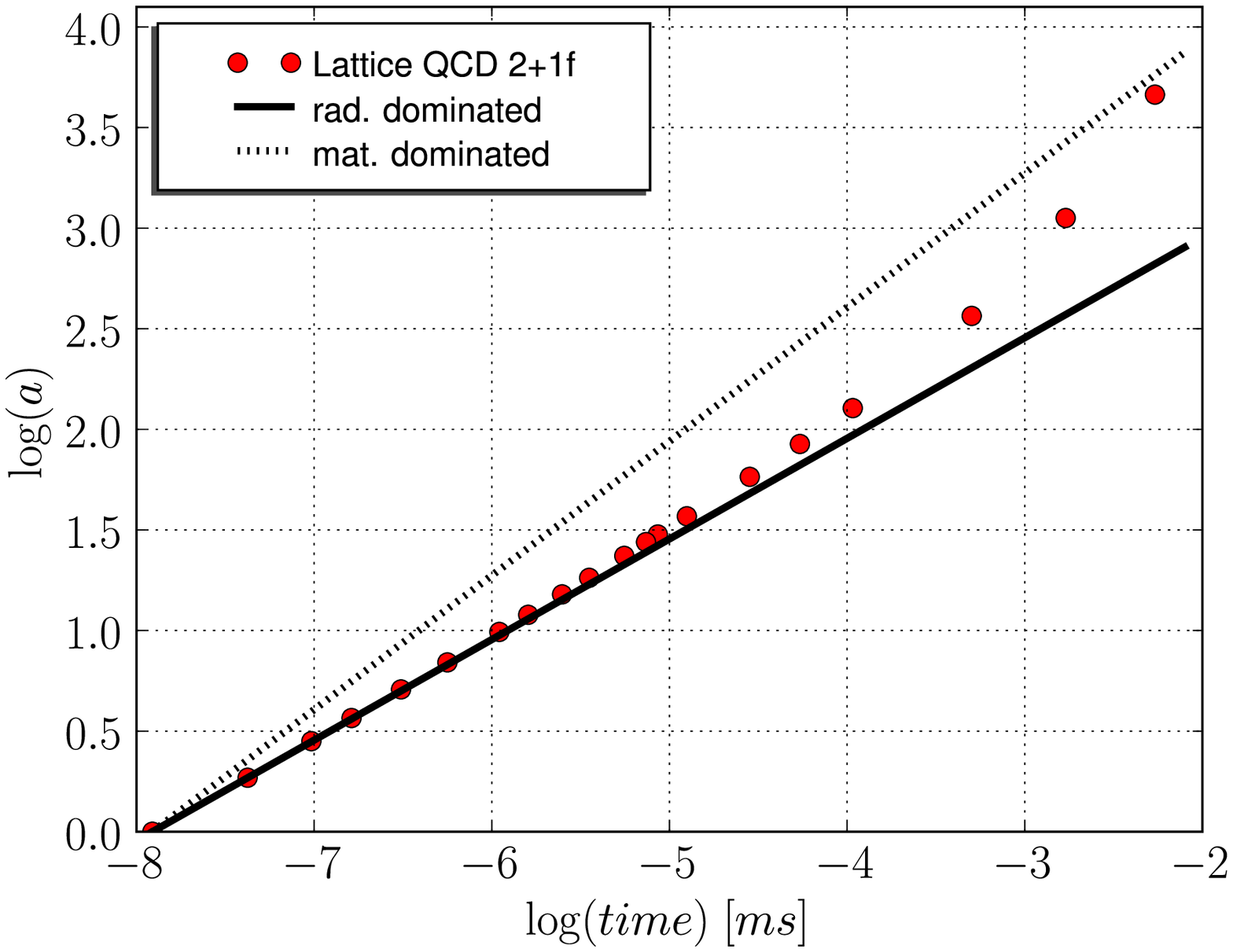}
     }
  }
  \caption{Time versus temperature (left) and time dependence of the
    scale factor (right) from Lattice QCD data. The behavior of $a(t)$
    for radiation (black line) and matter (black dashes) is shown. We
    set $t_{min} = t(T=719MeV)\equiv ( \frac{32\pi G}{3} \varepsilon(T=719MeV))^{-1/2} $ in $ms$ for $a(t_{min})=1.$}
  \label{fig2}
\end{figure}
The data for energy density $\varepsilon(T)$, pressure $p(T)$ and
trace anomaly $\varepsilon - 3p$ and entropy $s$ of
Ref.~\cite{Cheng:2007jq} are given in Table~\ref{tab1}. All the
analysis in this section is derived from this data. Besides the strange quark one can also include the effect of the charm quark as well as photons and leptons on the equation of state. These have important cosmological contributions as was shown in \cite{Laine:2006cp}. Recent references on lattice QCD at high temperature are \cite{Cheng:2007wu}\cite{Endrodi:2007tq}\cite{Miller:2006hr}.

\begin{table}[t]
\caption{\label{tab1}
Data from lattice QCD study with $N_t =6$~\cite{Cheng:2007jq}. Temperature in units of $(100 MeV)$, Energy density, pressure and trace anomaly in units of $(100 MeV)^4$, entropy density $s$ in units of $(100 MeV)^3$.}
\begin{center}
\begin{tabular}{|c|c|c|c|c|}
\hline  $T$ & $\varepsilon $ & $p$  & $\varepsilon - 3p$  & $s$\\
\hline 1.40 & 2.54  & 0.189  & 1.98	&1.96\\
\hline 1.45 & 6.24  & 0.315  & 5.30	&4.51\\
\hline 1.59 & 8.88  & 0.909   & 6.15	&6.17\\
\hline 1.74 & 21.4  & 2.45  & 14.0	&13.7\\
\hline 1.80 & 38.7  & 3.57   & 27.9	&23.5\\
\hline 1.86 & 58.7  & 5.35  & 42.7	&34.4\\
\hline 1.96 & 113  & 10.1   & 82.7	&62.9\\
\hline 2.03 & 170  & 15.3  & 124	&91.3\\
\hline 2.06 & 196  & 18.6  & 140	&104\\
\hline 2.13 & 238  & 26.6  & 158	&124\\
\hline 2.27 & 341  & 47.3   & 199	&171\\
\hline 2.40 & 449  & 74.1   & 226	&218\\
\hline 2.61 & 635  & 127  & 254	        &292\\
\hline 2.82 & 859  & 196  & 272	        &375\\
\hline 3.24 & 1520  & 398  & 326	&593\\
\hline 3.67 & 2520  & 714  & 377	&882\\
\hline 4.19 & 4350  & 1300  & 461	&1350\\
\hline 4.68 & 6800  & 2070  & 587	&1900\\
\hline 5.56 & 13800  & 4300  & 915	&3260\\
\hline 7.19 & 39300  & 12600  & 1480	&7220\\
\hline
\end{tabular}
\end{center}
\end{table}

We plot the energy density and pressure  in Figure~\ref{fig1}. For comparison, we plot the corresponding curves one
would expect for radiation behavior. While for the high temperature
regime we see, as expected, radiation like behavior, in the region at
and below the critical temperature $T_c$ ($\approx 200 MeV$) of the
deconfinement transition the behavior changes drastically. This change
in the behavior will also be relevant for cosmological observables as
we will see in the following.

For high temperature between $2.82$ $(100 MeV)$ and $7.19$ $(100 MeV)$ one can fit the data to a simple equation of state of the form:
\begin{equation}
\begin{array}{l}
 \varepsilon (T) = \alpha_{rad} T^4 + \dots \\ 
 p(T) = \sigma_{rad} T^4 + \dots  \\ 
 \end{array}
\label{eq:epspT}
\end{equation}
We find $\alpha_{rad} = 14.9702 \pm .009997$ and $\sigma_{rad} = 4.99115 \pm .004474$ using a least squares fit. The dots indicate terms constant and quadratic in temperature \cite{Cheng:2007jq}.

\begin{figure}[tp]

   \centerline{\hbox{
   \epsfxsize=3in
   \epsffile{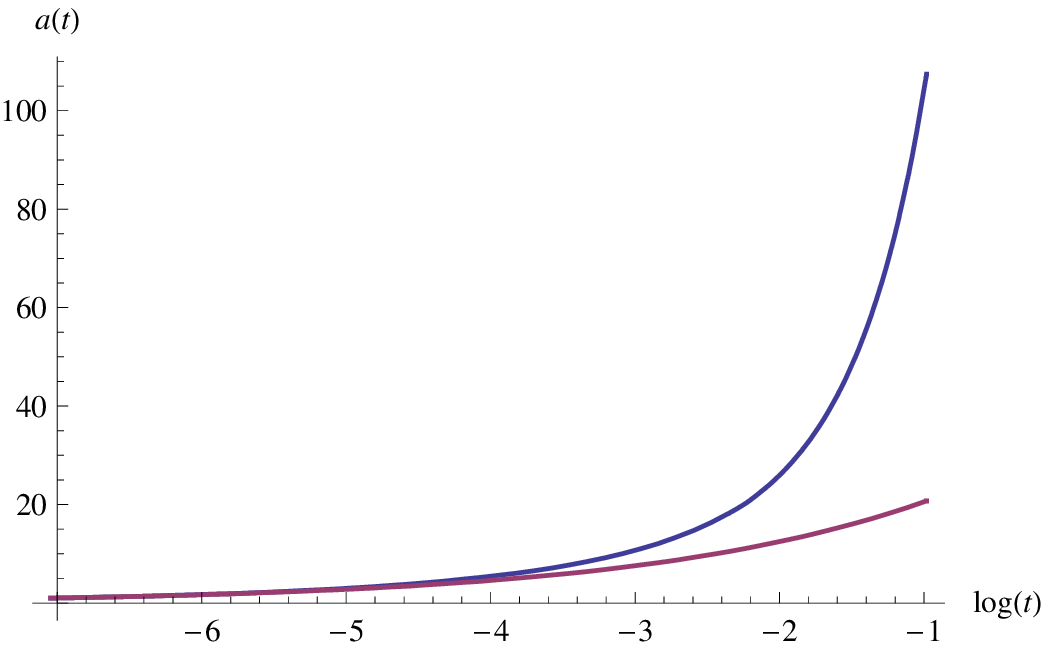}
   \epsfxsize=3in
   \epsffile{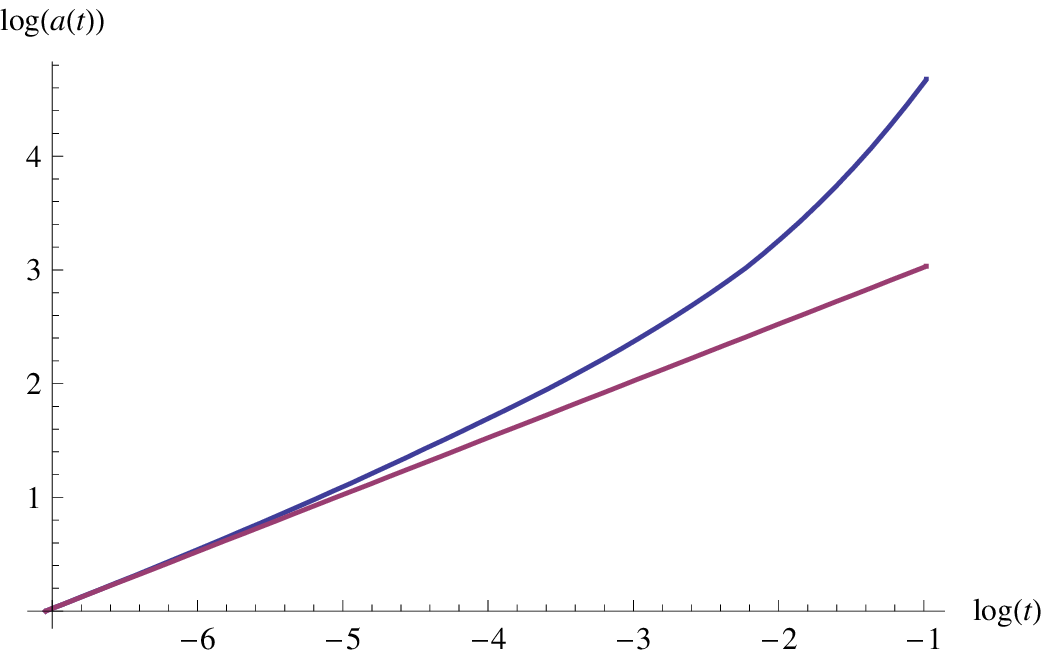}
     }
  }
  \caption{Scale factor and log of scale factor as function of log time from lattice equation of state. Lower curve is the result from radiation. A different fitting procedure using interpolation as opposed to Figure 2. This lead to a larger value for the scale factor. In this figure we set $\log t_{min} =\log t(T=719MeV)=-7.045967106$}

  \label{fig3}

\end{figure}

For lower temperatures the equation of state is very complex and use a
numerical procedure to determine $t(T)$, $T(t)$, $\varepsilon(T(t))$
and $a(t)$ using formulas~(\ref{eq:dTdt}), (\ref{eq:tT}) and (\ref{eq:at}). In
Figure~\ref{fig3} we plot $a$ and $\log a$ as a function of $\log t$.
The slope of $\frac{1}{2}$ at small times is indicative of expansion
due to radiation of the form $a(t) = t^{1/2}$ at high temperatures and
early times as discussed in section 1. At late times one see that the
scale factor dramatically increases and gives rise to characteristic
hockey stick shape pointing northeast.  This is reflective of the
shape of energy density and pressure plots as a function of
temperature which are hockey stick shapes at low temperature pointing
southwest.  Within this procedure we tried to avoid introducing new
systemic uncertainties and keep as close to the lattice data as
possible. In
Figure~\ref{fig2} we show time vs.~temperature (left) and the time
dependence of the scale factor (right). We see that in the confinement
region, i.e.~ for $T$ less than 200 $MeV$ or $\log(t)$ greater than
$-5$, the behavior changes.  In plot of the right hand side of
Figure~\ref{fig2} we also show the behavior of $a(t)$ for radiation
and matter.  While for times before the phase transition the lattice
data matches with the radiation behavior very well, for times
corresponding to temperatures above $T_c$ the behavior of the lattice
data changes towards matter dominated behavior. We remark that lattice
studies show that the QCD phase transition at its physical values is
actually a cross-over phase transition. Therefore, the change in the
scale factor we observe is rather moderate and is in contrast to a
scenario one would expect, e.g., from a first order phase transition.

Besides lattice QCD there are other approaches to the low temperature equation of state. In the following sections we compare the prediction of some of these approaches for $\varepsilon(T)$, $p(T)$ and the scale factor $a(t)$ to the results from lattice QCD. In particular we shall discuss how well various models can describe the characteristic shape of the $\log a$ versus $t$ curves coming from lattice QCD.

\section{Hadronic resonance gas model}

In the framework the Hadronic Resonance Gas model (HRG), QCD in the
confinement phase is treated as an non-interacting gas of fermions and
bosons~\cite{Karsch:2003vd,Karsch:2003zq,SakthiMurugesan:1990pe,Tawfik:2004sw}.
The fermions and bosons in this model are the hadronic resonances of
QCD, namely mesons and baryons. The idea of the HRG model is to
implicitly account for the strong interaction in the confinement phase
by looking at the hadronic resonances only since these are basically
the relevant degrees of freedom in that phase. The HRG model is
expected to give a good description of thermodynamic quantities in the
transition region from high to low
temperature~\cite{BraunMunzinger:2003zd}.

The partition function of the HRG model is given by a sum of one
particle partition functions,
\begin{equation}
  \label{eq:onepartZ}
  \log Z(T,V) = \sum_i \log Z^1_i(T,V)= \sum_i \frac{V g_i}{2\pi^2}
  \int_0^\infty dp \, p^2 \eta_i \log(1+\eta_i \mathrm{e}^{-\beta E_i}).
\end{equation}
The HRG model includes hadronic masses $m_i$ and degeneracies $d_i$ in
a low energy statistical model. The equation of state is given by:
\[
\begin{array}{l}
 \varepsilon (T) = \sum\nolimits_{m_i  } {\frac{{d_i }}{{2\pi ^2 }}} \sum\limits_{k = 1}^\infty  {( - \eta _i )^{k + 1} m_i^4 } ((\frac{T}{{km_i }})K_1 (km_i /T) + 3(\frac{T}{{km_i }})^2 K_2 (km_i /T)), \\ 
 p(T) = \frac{1}{3}\sum\nolimits_{m_i  } {\frac{{d_i }}{{2\pi ^2 }}} \sum\limits_{k = 1}^\infty  {( - \eta _i )^{k + 1} m_i^4 } 3(\frac{T}{{km_i }})^2 K_2 (km_i /T).
 \end{array}
\]
Here $\eta_i = -1$ for bosons and $\eta_i = 1$ for fermions and
$K_1(z)$ and $K_2(z)$ are modified Bessel functions. For practical
reasons we performed the sum over $m_i$ up to $m_{max} = 2.5 GeV$.

\begin{figure}[tbp]
   \centerline{\hbox{
   \epsfxsize=3.0in
   \epsffile{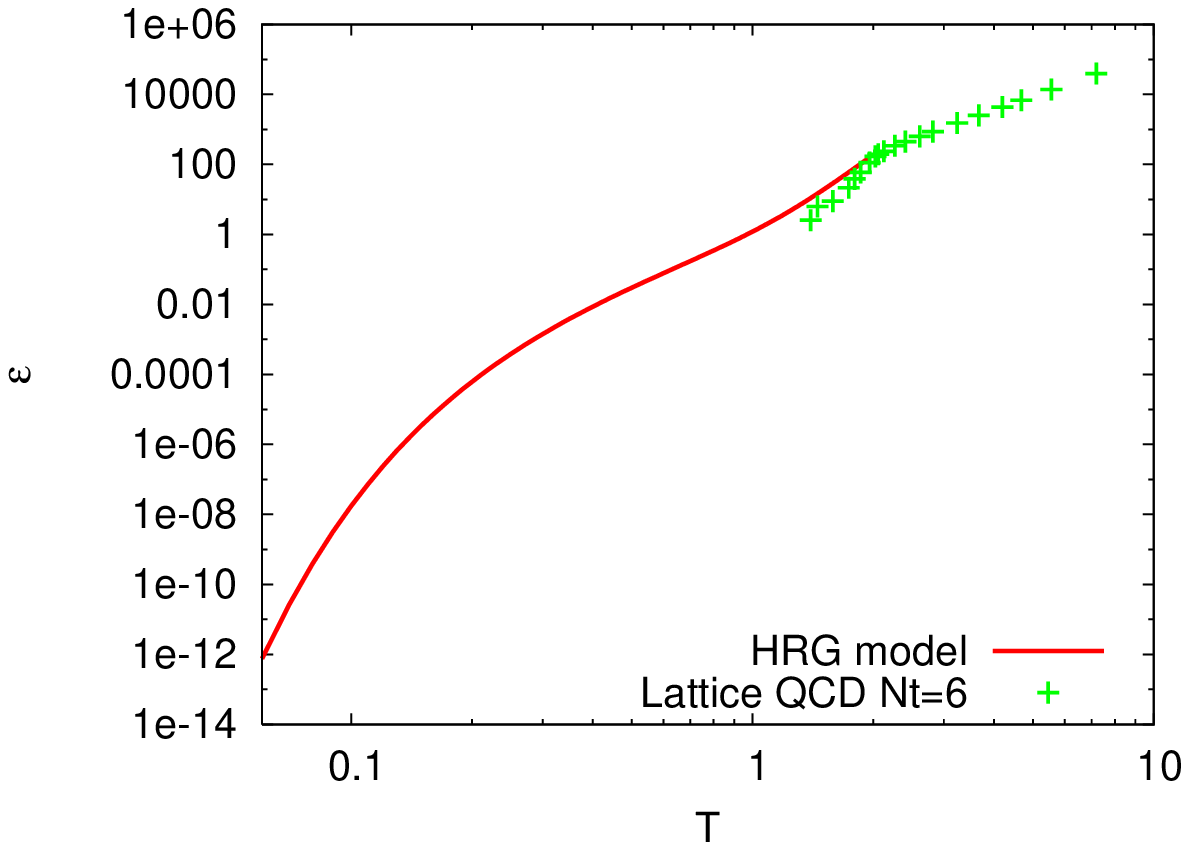}
   \epsfxsize=3.0in
   \epsffile{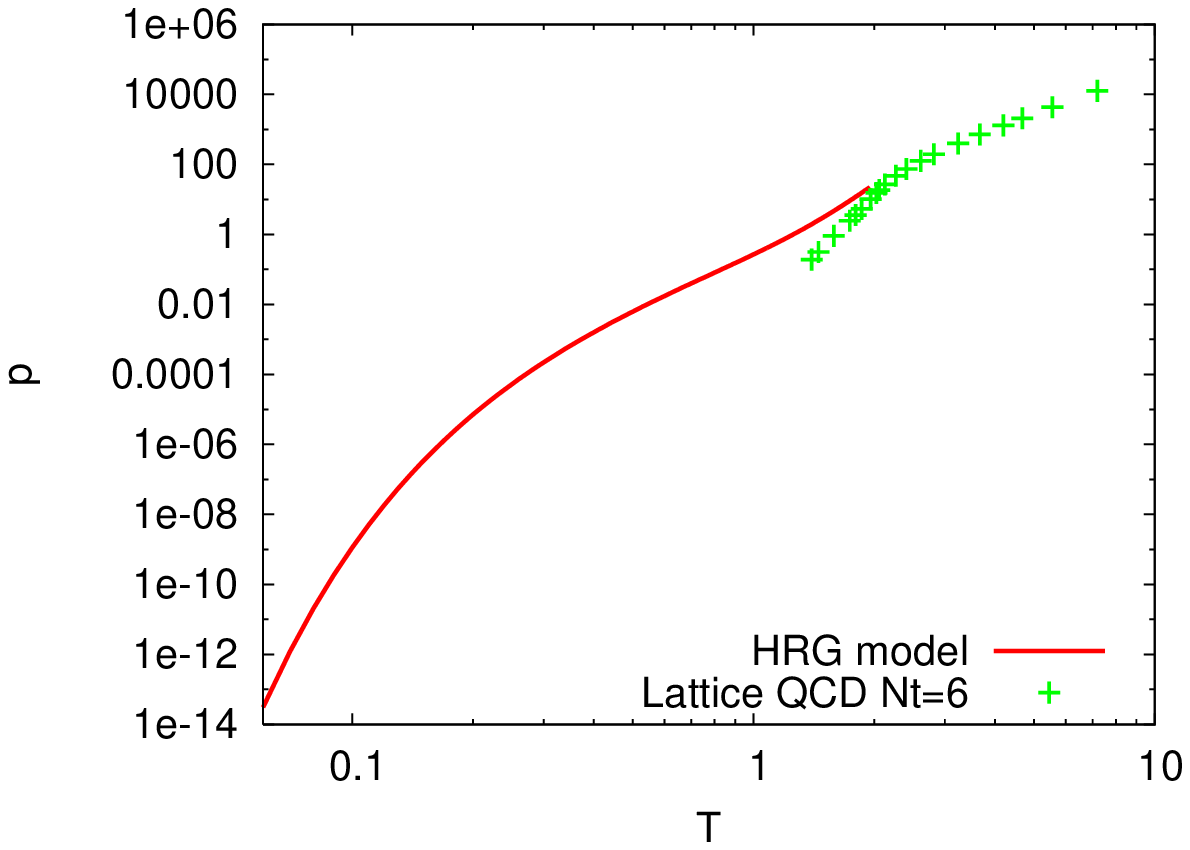}
     }
  }
  \caption{Energy density and pressure from the HRG  equation of state. The crosses are lattice QCD data with $N_t=6$}
  \label{fig4}
\end{figure}

\begin{figure}[tbp]
   \centerline{\hbox{
   \epsfxsize=3.0in
   \epsffile{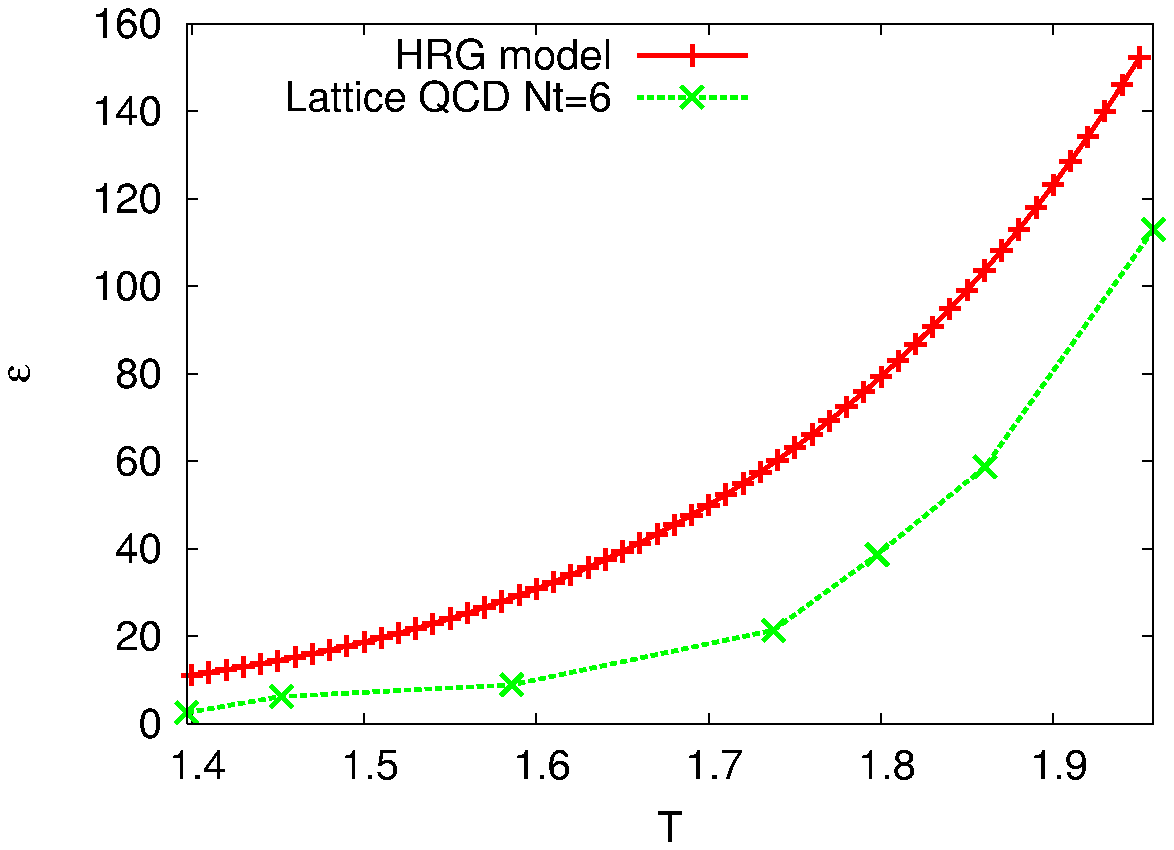}
   \epsfxsize=3.0in
   \epsffile{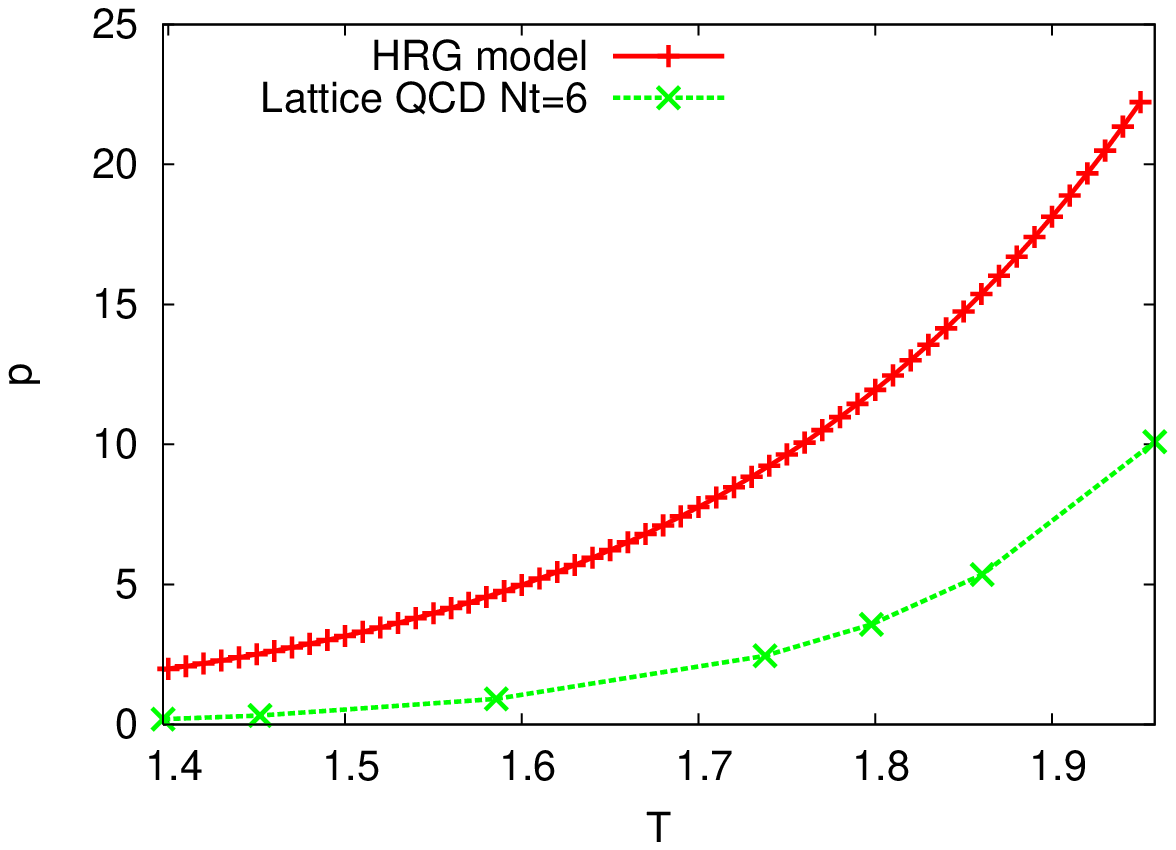}
     }
  }
  \caption{Energy density and pressure from the HRG equation of state over the range of temperatures covered by the lattice QCD calculation. The crosses are lattice QCD data with $N_t=6$.}
  \label{fig5}
\end{figure}

\begin{figure}[tbp]
   \centerline{\hbox{
   \epsfxsize=3.0in
   \epsffile{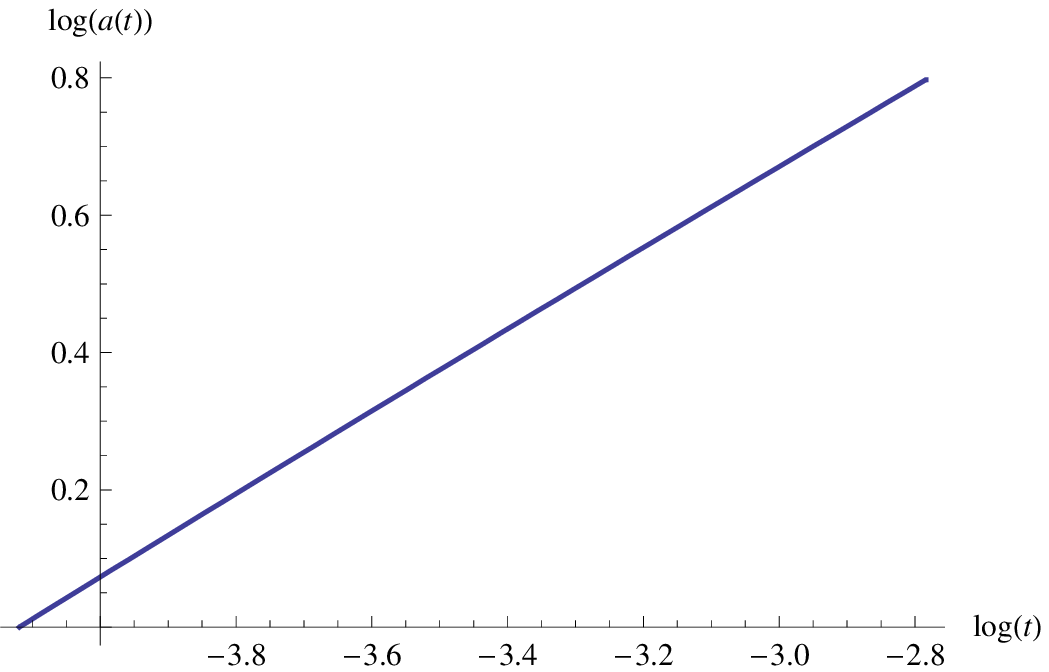}
   \epsfxsize=3.0in
   \epsffile{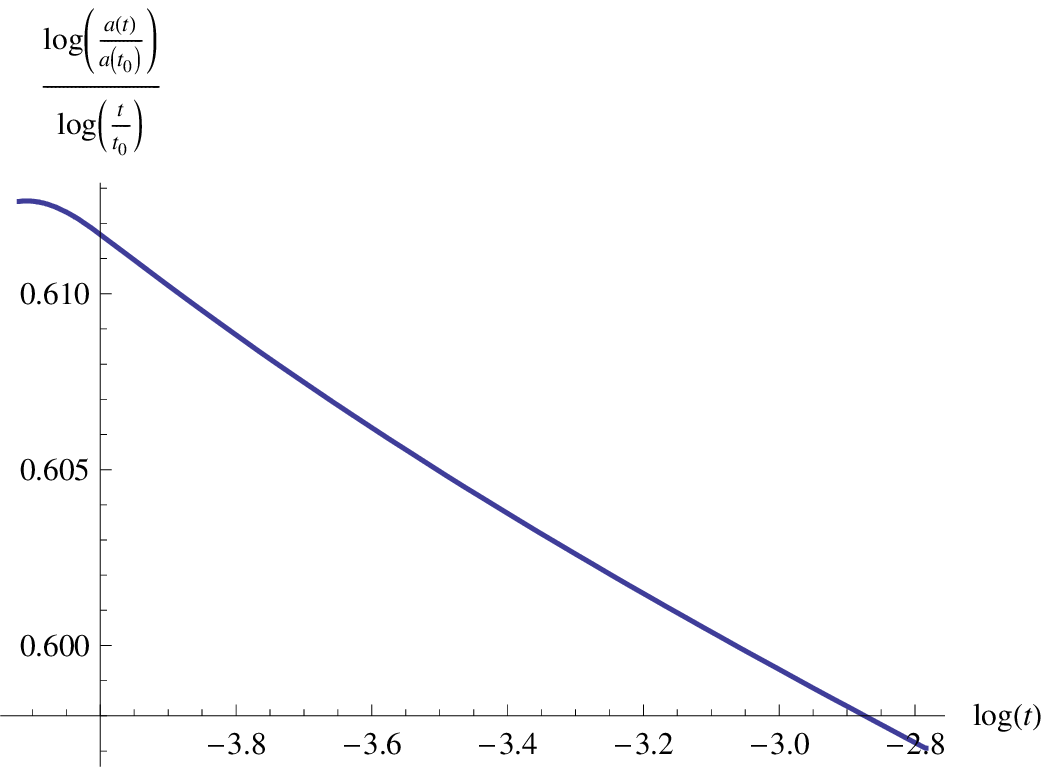}
     }
  }
  \caption{HRG model $\log a(t)$ as a function of $\log t $. The time range corresponds to temperatures from 1.39 to 1.95 in units of 100 MeV. The slope is approximately .605 which is approaching the matter dominated value of .666. Here $t_0$ is .0162553.}
  \label{fig6}
\end{figure}

We plot the energy density and pressure as a function of temperature
in Figure~\ref{fig4}. Over the temperature range of the the lattice
QCD data the HRG model gives values of the energy density and pressure
that exceed the lattice QCD data as shown in
Figure~\ref{fig5}~\cite{Cheng:2007jq}.  If one derives the scale
factor between 1.39 and 1.95 in units of 100 MeV one sees a power
expansion with coefficient .605 which is less than the matter
dominated result of $2/3$ as shown in Figure~\ref{fig6}.  This is
reasonable as this region of temperature is intermediate between the
matter dominated phase with expansion exponent $2/3$ and radiation
dominated phase with expansion coefficient $1/2$. The HRG model
equation of state is expected to be valid at low temperatures with matter
dominated expansion. As the temperature is increased the value of the
expansion exponent drops as one enters the regime best described by
high temperature lattice QCD. Ultimately in the full QCD theory the
expansion exponent drops as one reaches deconfinement where the theory
is described by a quark-gluon plasma.

\section{Hagedorn model}

This model is similar to the HRG model except the degeneracy factors take a string type dependence on mass \cite{Hagedorn:1965st,Huang:1970iq}. The energy density and pressure are given by:
\[
\begin{array}{l}
 \varepsilon (T) = \int {dm\rho (m)\int {\frac{{d^3 k}}{{(2\pi )^3 }}} } \frac{1}{{e^{E(k,m)/T}  + 1}}E(k,m), \\ 
 p(T) = \frac{1}{3}\int {dm\rho (m)\int {\frac{{d^3 k}}{{(2\pi )^3 }}} } \frac{1}{{e^{E(k,m)/T}  + 1}}\frac{{k^2 }}{{E(k,m)}}
 \end{array}
\]
where $E(k,m) = \sqrt {k^2  + m^2 }$. The degeneracy function $\rho(m)$ is given by:
\[
\rho(m) = c m^{-\gamma} \exp (\frac{m}{T_H}).
\]
for large mass $m$ and some exponent $\gamma$.  These formulas have simple generalizations to nonzero chemical potential, although we restrict ourselves to zero chemical potential in this paper. The integrals over $m$ and momentum can be done using the methods of Carlitz \cite{Carlitz:1972uf}. For $\gamma = 5/2$ one has the equation of state:
\[
\begin{array}{l}
 p(T) = \alpha_0 (\frac{T}{{T_H }})^{5/2} \Gamma (\frac{{m_0 (T_H  - T)}}{{T_H T}},0), \\ 
 \varepsilon (T) = \frac{\alpha_0 }{{T_H }}T^2 \partial _T ((\frac{T}{{T_H }})^{3/2} \Gamma (\frac{{m_0 (T_H  - T)}}{{T_H T}},0))
 \end{array}
\]
where $\Gamma(x,\frac{5}{2}-\gamma) = \int^{\infty}_x e^{-y} y^{3/2-\gamma} dy$. For $\gamma = 5/2$ this is the exponential integral $E_1(x)= \int^{\infty}_x e^{-y} y^{-1} dy$.

\begin{figure}[tbp]
   \centerline{\hbox{
   \epsfxsize=3.0in
   \epsffile{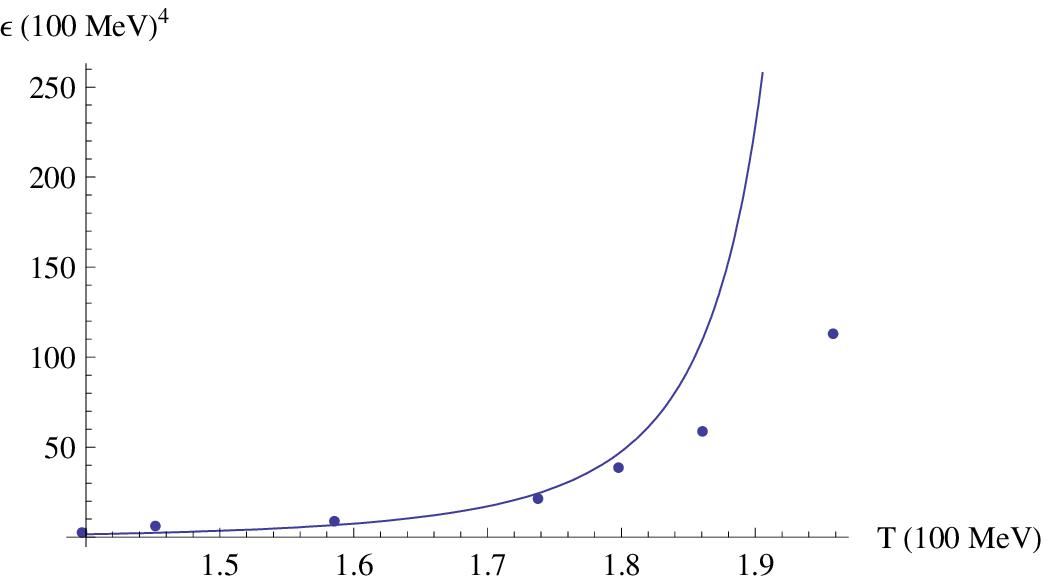}
   \epsfxsize=3.0in
   \epsffile{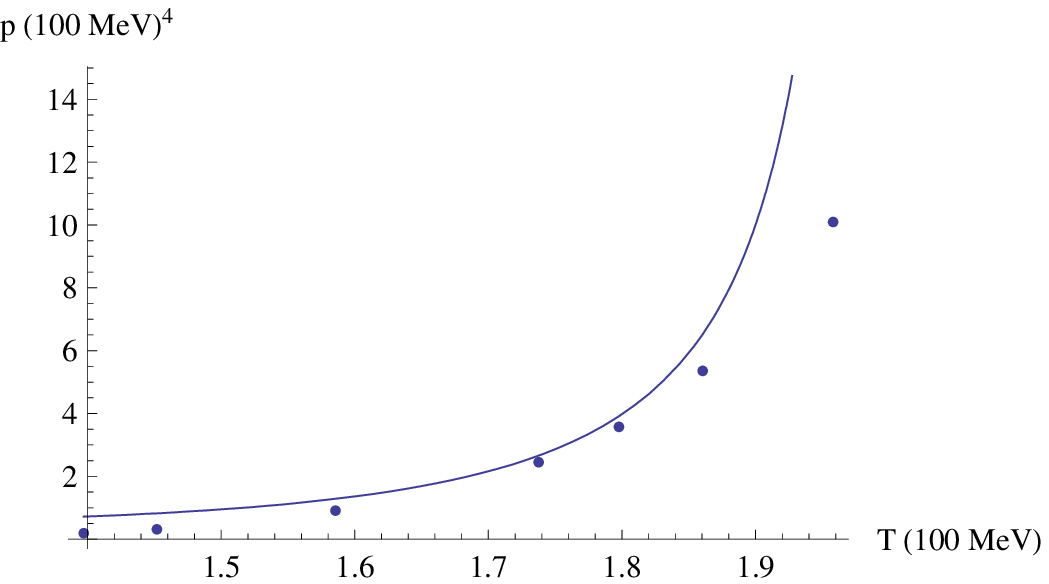}
     }
  }
  \caption{Energy density and pressure from the Hagedorn equation of state over the range of temperatures covered by the lattice QCD calculation. The dots are lattice QCD data with $N_t=6$.}
  \label{fig7}
\end{figure}

Pressure and energy density are are plotted in Figure~\ref{fig7} for $T_H = 200
MeV$, $\alpha_0 = 1.96665 (100 MeV)^4$ and $m_0 = 6.10776 (100 MeV)$
alongside the lattice data. The energy density displays a limiting
temperature so we study the equation of state of the Hagedorn model as
a model of low energy QCD below $T_H$ only. The approach of using a
string-like model with a limiting temperature to describe the strong
interactions has a long history. A modern perspective on the approach
is given in \cite{Castorina:2007eb,Harmark:2006ta,KalyanaRama:1998cb}.
In $TeV$ scale gravity a $TeV$ Hagedorn temperature is possible if the
string scale turns out to be at a few $TeV$\cite{Antoniadis:1999fj}. A
Hagedorn type cosmology in the early Universe is proposed as
alternative to inflation in with the Hagedorn temperature at two
orders of magnitude below the Planck scale \cite{Nayeri:2005ck}. A
nice discussion at the popular level of the concept of limiting
temperature or absolute hot is given in \cite{Tyson}.

\begin{figure}[tbp]
   \centerline{\hbox{
   \epsfxsize=3.0in
   \epsffile{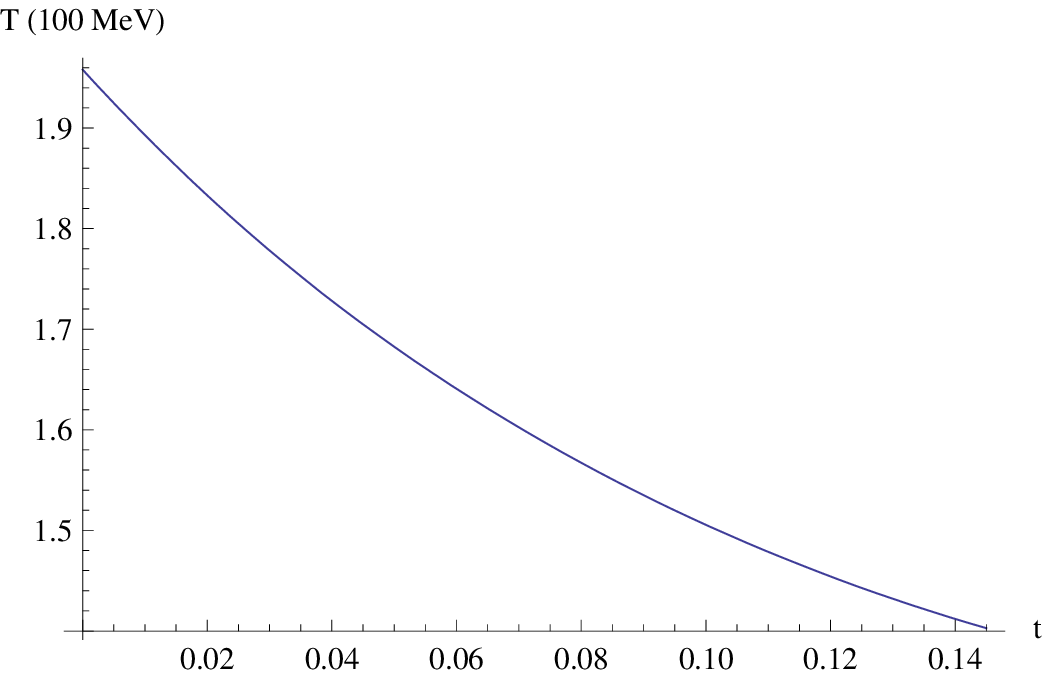}
   \epsfxsize=3.0in
   \epsffile{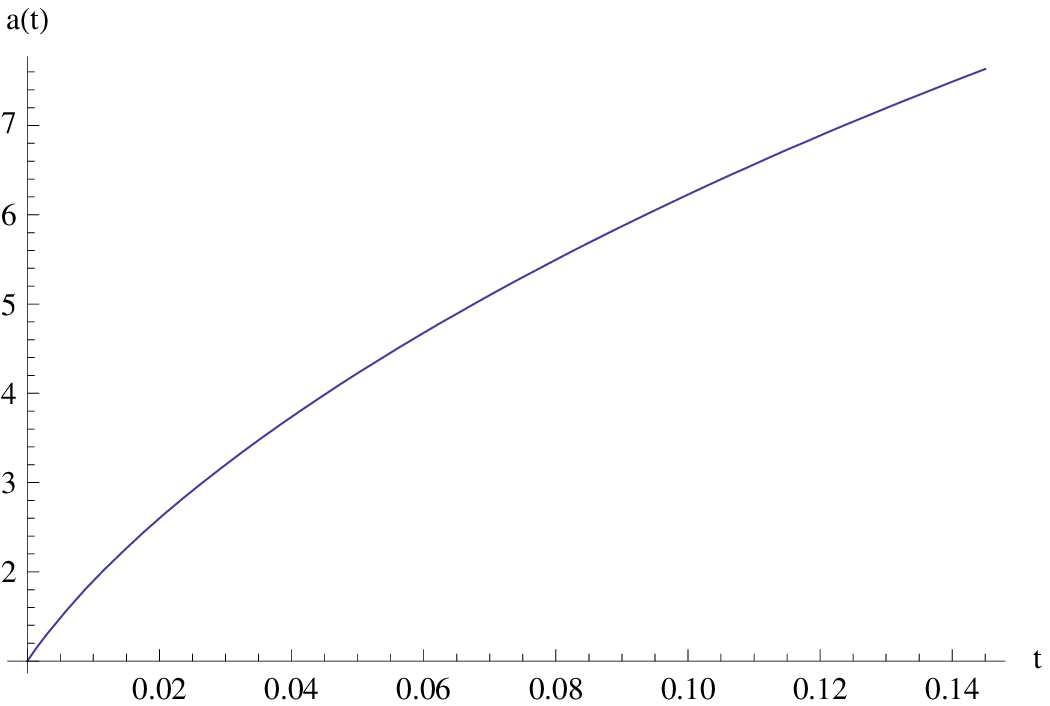}
     }
  }
  \caption{Temperature and scale factor as a function of time from the Hagedorn equation of state.}
  \label{fig8}
\end{figure}

Given the equation of state we can use the methods of the previous
section to determine the temperature a function of time as well as the
scale factor $a(t)$ which are shown in Figure~\ref{fig8}. Note from
Figure~\ref{fig8} the temperature is less than $T_H$ for all times in
the Hagedorn model of cosmology. This shows the limiting
temperature nature of the Hagedorn cosmology. Another feature of the Hagedorn cosmology is that $a(t)$ in
Figure~\ref{fig8} is more complex than a power law as was shown
in~\cite{Huang:1970iq}. The Hagedorn cosmology is not considered a
leading cosmological theory at this time mainly because temperatures
higher than $T_H$ can be observed in the early Universe. However for
theories where the Hagedorn temperature is at the TeV scale or at the
Planck length the Hagedorn cosmology is still of interest.

\section{AdS/CFT model of low energy QCD}

Determining the  time dependence of the scale factor $a(t)$ is not one of the strong points of the AdS/CFT correspondence. This is because the only gravity in the correspondence is induced in the AdS five dimensional space and there is no dynamical gravity on the CFT side which one wants to identify with a QCD-like theory. Thus questions like what happens when a glue-ball falls into a black hole or even how a proton is attracted to the earth can't be straightforwardly addressed within $AdS_5/CFT_4$. One has to use another approach with gravity and matter on the same side of the correspondence.  Nevertheless, recently Gubser and Nellore \cite{Gubser:2008ny}  have determined a low energy equation of state for a QCD-like theory using $AdS^5/ CFT_4$ and these can be used to calculate the back reaction on Einstein's equation to leading order.

We first consider $AdS$ gravity without dilaton or with dilaton equal to zero, where the entropy and mass can be determined analytically. We then consider refinements from including the dilaton and a numerical treatment similar to \cite{Gubser:2008ny} to determine the entropy and mass. Then using the conjectured duality between the AdS black hole and CFT theories we obtain the entropy and energy density of the dual gauge theory and it's equation of state. Finally one can use this entropy and energy density to determine the expansion\ factor $a(t)$ associated with the dual gauge theory by using Einstein's equations on the the dual gauge side of the correspondence.

\subsection{$AdS^5$ gravity without the dilaton}

We work within the ansatz for the five dimensional metric given by:
\[
ds^2  =  c^2 (r)( - dt^2 )+ b^2 (r)dr^2  + \tilde a^2 (r)d\Omega _3^2.
\]
We use a tilde to differentiate the dual variable $\tilde a$ from the scale factor $a$ that occurs in the physical four dimensional metric. We denote the metric for the unit three sphere by $d\Omega _3^2$. 

The equations of motion within this ansatz follow from the Lagrangian:
\[
L = \left(6\frac{{\tilde a'^2 }}{{\tilde a^2 b}} + 6\frac{{\tilde a'}}{{\tilde a}}\frac{{c'}}{{bc}} + 6\frac{b}{{\tilde a^2 }} + \tilde \lambda b - \frac{{\phi '^2 }}{{2b}} - V(\phi )b\right)\tilde a^3 c
\]
which comes from the Einstein-Hilbert Lagrangian:
\[
\sqrt{-g}(R + \tilde \lambda - \frac{1}{2} (\partial \phi)^2 - V(\phi)).
\]
Where $-\tilde \lambda $ denotes the cosmological constant of the dual $AdS$ space and $V(\phi)$ is a potential of a scalar field $\phi$. We impose the gauge condition:
\[
cb = 1.
\]
In this gauge the equations of motion become:
\[
\begin{array}{l}
 12\alpha '' + 18\alpha '^2  + 24\alpha '\beta ' + 6\beta '' + 12\beta '^2  - 6e^{2\alpha  - 2\beta }  - 3\tilde \lambda e^{- 2\beta }  + \frac{3{\phi '^2 }}{2} + 3V(\phi )e^{- 2\beta }  = 0 \\ 
 6\alpha '' + 6\alpha '^2  + \phi '^2  = 0 \\ 
 \phi '' + 3\phi '\alpha ' + 2\phi '\beta ' - \frac{{dV(\phi )}}{{d\phi }}e^{- 2\beta }  = 0 \\ 
 \end{array}
\]
where we have defined $\alpha = \log {\tilde a}$ and $\beta = \log c$.

\begin{figure}[tbp]
   \centerline{\hbox{
   \epsfxsize=3.0in
   \epsffile{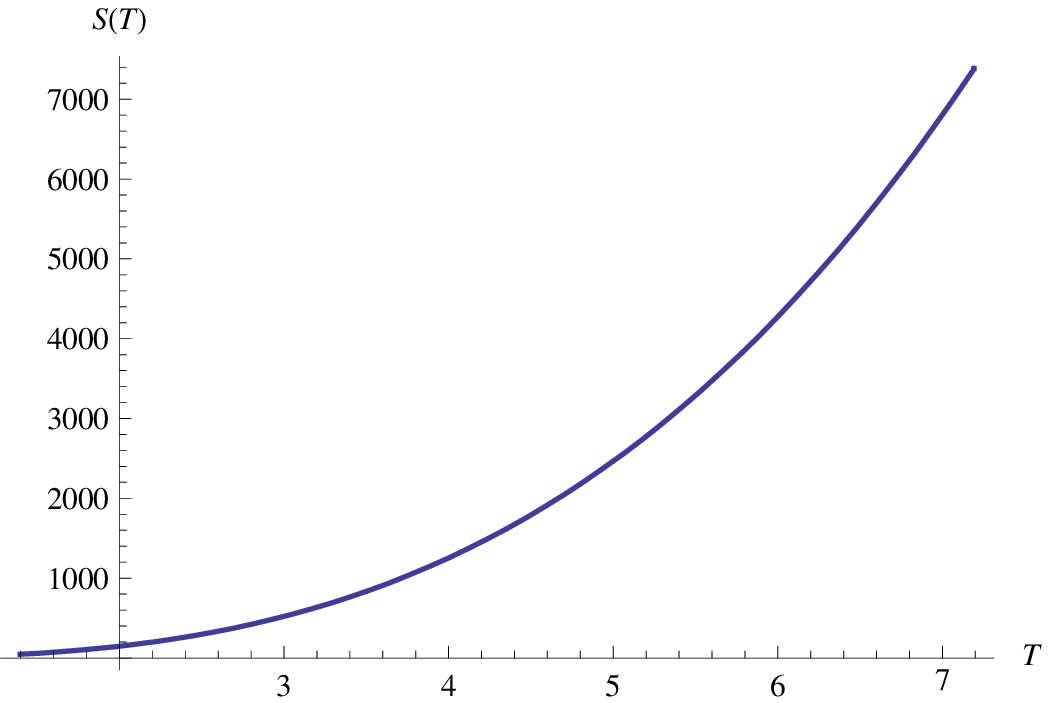}
   \epsfxsize=3.0in
   \epsffile{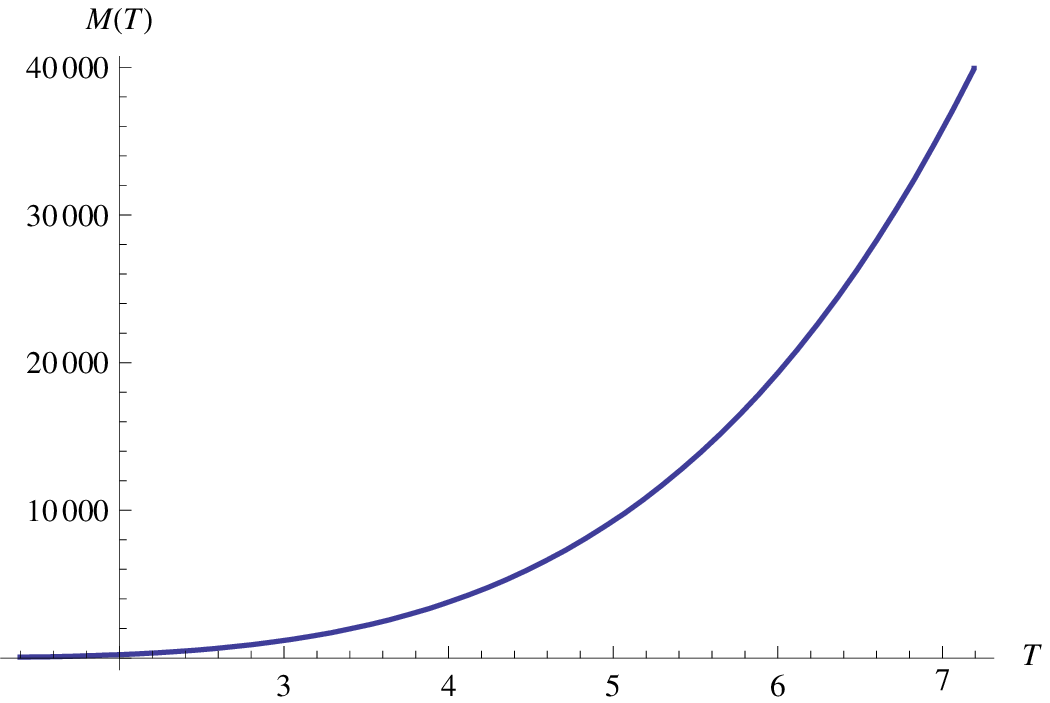}
     }
  }
  \caption{Entropy and mass (energy density) of the $AdS^5$ black hole
    as function of temperature without the dilaton.}
  \label{fig9}
\end{figure}

The equations have the solution in vacuum $\phi = 0$ and $V(\phi)=0$ given by:
\[
\begin{array}{l}
 \tilde a = r \\ 
 b = (1 -\frac{\mu}{r^2}  + \frac{\tilde \lambda }{12}r^2 )^{-1/2} \\ 
 c = (1 -\frac{\mu}{r^2}  + \frac{\tilde \lambda }{12}r^2 )^{1/2} \\
 \end{array}
\]
In this solution $\mu$ is a constant parameter which turns out to be be proportional to the mass of the black hole. The horizon is determined by the largest solution to $c(r_+)=0$ and is given by:
\[
r_ +   = \mu^{1/2}\frac{\sqrt{2}}{\sqrt {{{1 + \frac{{\sqrt {3 + \mu\tilde \lambda } }}{\sqrt 3}}}}} .
\]
The entropy can be computed from the formula:
\[
S = \frac{{\omega_3}}{4}\tilde a^3 \tilde m_P^3 |_{r = r_+}.
\]
where $\omega_3 = 2\pi^2$ is the volume of a unit three sphere and $\tilde m_P$ is the five dimensional Planck mass in the dual space. The temperature of the black hole solution is determined from:
\[
T = \frac{1}{{4\pi }}\frac{{\tilde a^2 }}{{cb}}(\frac{c^2}{\tilde a^2 })'|_{r = r_+}.
\]
One can define a mass formula similar to that of Poisson and Israel \cite{Poisson:1990eh}  and Fischler, Morgan and Polchinski \cite{Fischler:1990pk} for spherically symmetric gravity. It is given by:
\[
M = \frac{3\omega_3}{16\pi} \tilde m_P^3\tilde a^2 (1 - \frac{{\tilde a'^2 }}{{b^2 }} + \frac{{\tilde \lambda }}{{12}}\tilde a^2 )|_{r = \infty}.
\]
Applying these formula to the above solution the entropy is given by:
\[
S  = \frac{{\omega_3 }}{4}\mu ^{3/2}\tilde m_P^3 \left( {\frac{2}{{1 + \frac{{\sqrt {3 + \mu \tilde \lambda } }}{{\sqrt 3 }}}}} \right)^{3/2} 
\]
and temperature is 
\[
T  = \frac{1}{2 {\pi \mu ^{1/2} }}\sqrt {\frac{{1 + \frac{{\sqrt {3 + \mu \tilde \lambda } }}{{\sqrt 3 }}}}{2}} \frac{{\sqrt {3 + \mu \tilde \lambda } }}{{\sqrt 3 }}.
\]
The mass formula simply reduces to:
\[
M = \frac{3 \omega_3}{16 \pi } \mu \tilde m_P^3.
\]
This solution satisfies:
\[
T = (\frac {\partial S}{\partial M})^{-1}
\]
in analogy with classical thermodynamics \cite{Pidokrajt,Louko:1996jd,Quevedo:2008xn}. Expanding the expressions for entropy and mass as a function of temperature for large mass one finds:
\[
S \tilde m_P^3 = \frac{1}{2} \pi^5 (\frac{12}{\tilde \lambda})^3 \tilde m_P^6 T^3 + \dots
\]
and 
\[
M \tilde m_P^3 = \frac{3}{8} \pi^5 (\frac{12}{\tilde \lambda})^3 \tilde m_P^6 T^4 + \dots
\]
Matching these expressions to the lattice QCD data for entropy and
energy density at high temperatures from formula~(\ref{eq:epspT}) one
finds that
\[
\tilde \lambda \approx 23.661 \tilde m_P^2.
\]
The entropy and mass of the $AdS^5$ black hole solution without dilaton are plotted in Figure~\ref{fig9}.

\begin{table}[t]
\caption{Quantities used to compare $AdS^5$ gravity and finite temperature QCD cosmology.}
\label{tab2}
\begin{center}
\begin{tabular}{|c|c|c|}
\hline    Quantity & $AdS^5$ gravity & QCD cosmology\\
\hline entropy & $S \tilde m_P^3$  & $s$  \\
 energy density& $M \tilde m_P^3$ & $\varepsilon$ \\
 temperature& $T_{Hawking}$  & $T$\\
 fields & $\tilde a(r), b(r), c(r), \phi(r)$ & $a(t)$\\
 fundamental constants & $\tilde m_P$, $\tilde \lambda$ & $M_P$, $\lambda$\\
\hline\end{tabular}
\end{center}
\end{table}

\begin{figure}[tbp]
   \centerline{\hbox{
   \epsfxsize=3.0in
   \epsffile{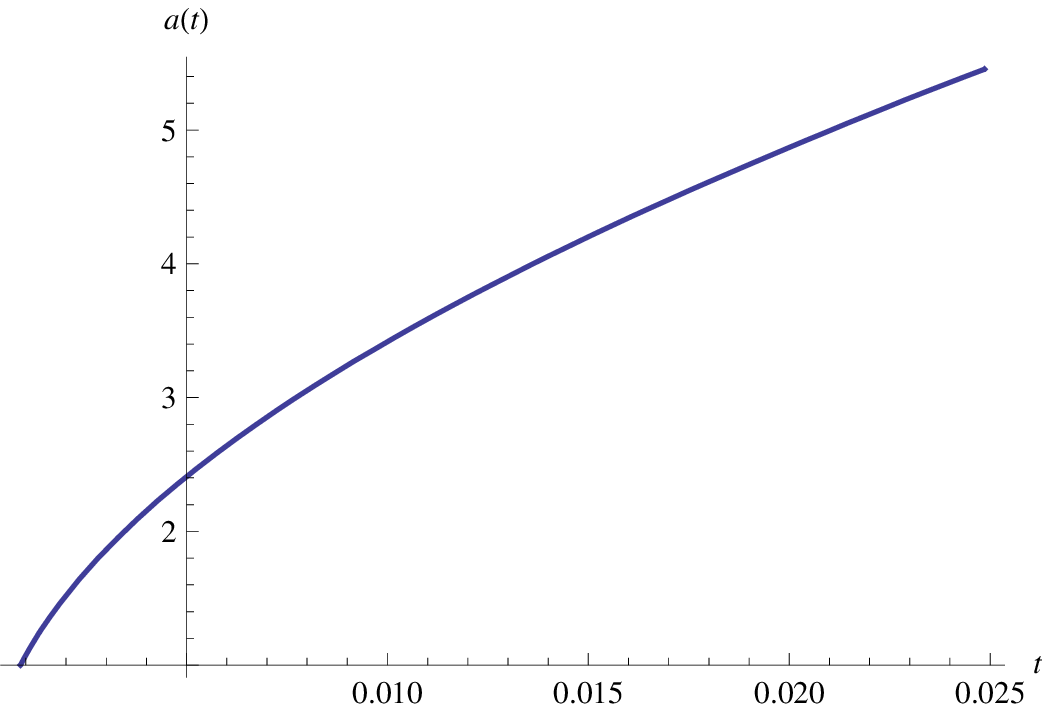}
   \epsfxsize=3.0in
   \epsffile{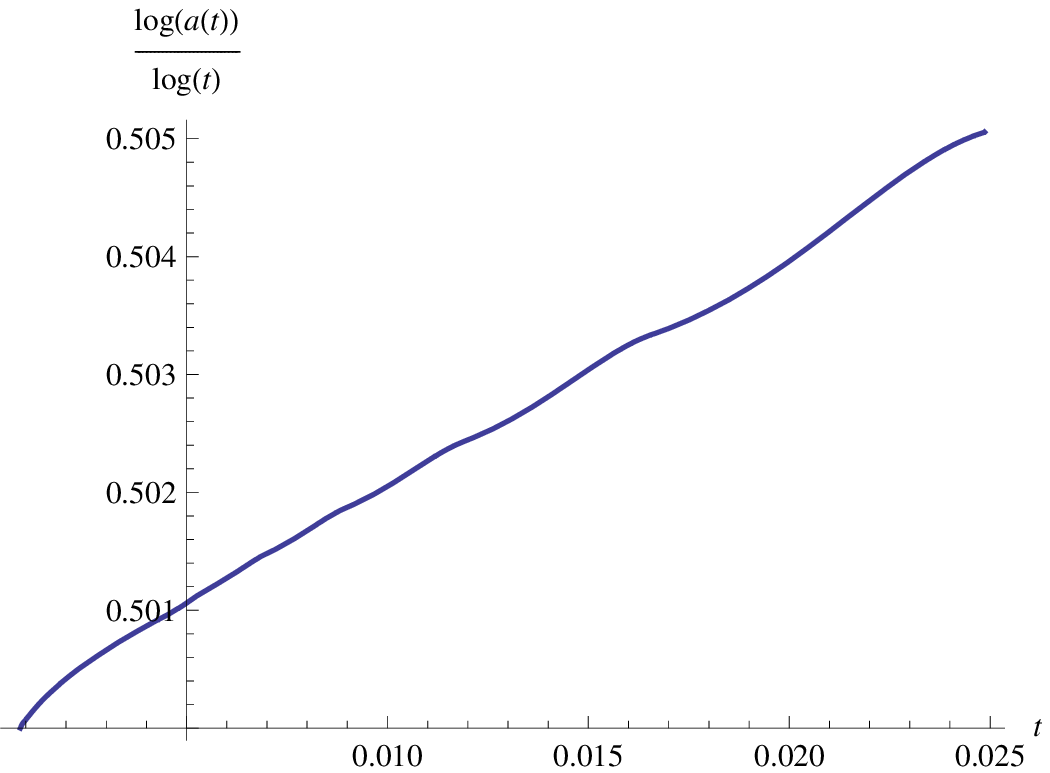}
     }
  }
  \caption{Time dependence of radius $a(t)$ determined from dual $AdS^5$  black hole solution without the dilaton.}
  \label{fig10}
\end{figure}

Rewriting the equation~(\ref{eq:dTdt}) that determines the time dependence of the temperature in terms of entropy and energy density we have:
\begin{equation}
\frac{{dT}}{{dt}} =  - 3(T s)\sqrt {\frac{{ \varepsilon (T)}}{3M_P^2}} (\frac{{d\varepsilon (T)}}{{dT}})^{ - 1} .
\label{eq:dTdts}
\end{equation}
Then using the correspondence from Table~\ref{tab2} one can determine
the radius of the 4d universe $a(t)$ from~(\ref{eq:dTdts}). This is
plotted in Figure~\ref{fig10}. One can see from the plot on right hand
side of Figure~\ref{fig10} that over the temperature range covered by
the lattice QCD data one never leaves the radiation regime and the
scale factor is well approximated by $a(t) = t^{1/2}$.

In this AdS approach it is is important not to confuse the scale
factor $a(t)$ with the dual radius $\tilde a(r)$. One uses the black
hole solution including $\tilde a(r)$ to determine the entropy,
temperature and energy density and then uses~(\ref{eq:dTdts}) to
determine the scale factor $a(t)$ through the correspondence in
Table~\ref{tab2}.

\subsection{ $AdS^5$ gravity with nonzero dilaton}

Because the vacuum solution to $AdS^5$ gravity does not match with the lattice QCD data at low temperature and stays in the radiation regime over the temperature range of lattice QCD, one looks for non vacuum solutions that can mimic the lattice QCD equation of state. In \cite{Gubser:2008ny} the potential 
\begin{equation}
V(\phi) = \tilde \lambda (1-\cosh(2.057 \phi) + \tilde \lambda \frac{0.606}{12} \phi^2
\end{equation}
was used to describe a $AdS^5$ black hole with nonzero dilaton field. For nonzero $\phi$ and $V(\phi)$ one can solve the equations of motion numerically. In \cite{Gubser:2008ny} it was shown that the speed of sound associated with the potential (5.2) closely approximates the speed of sound from lattice QCD.

For nonzero dilaton it is convenient to replace the coordinate $r$ by $1/z$. Then the asymptotic region $r = \infty$ corresponds to $z=0$. In the coordinate $z$, the five dimensional metric ansatz is taken to be:
\[
ds^2  =  c^2 (z)( - dt^2 )+ b^2 (z)dz^2  + \tilde a^2 (z)d\Omega _3^2 .
\]
It is convenient to choose the gauge:
\[
c b = \tilde a^2.
\]
In this gauge the equations of motion become:
\begin{equation}
\begin{array}{l}
 12\alpha '' + 6\alpha '^2  + 24\alpha '\beta ' + 6\beta '' + 12\beta '^2  - 18e^{2\alpha  - 2\beta }  - 5\tilde \lambda e^{4\alpha - 2\beta }  + \frac{{\phi '^2 }}{2} + 5 V(\phi )e^{4\alpha - 2\beta }  = 0 \\ 
 6\alpha '' - 6\alpha '^2  + \phi '^2  = 0 \\ 
 \phi '' + \phi '\alpha ' + 2\phi '\beta ' - \frac{{dV(\phi )}}{{d\phi }}e^{4 \alpha- 2\beta }  = 0 \\ 
 \end{array}
\end{equation}
where the prime refers to the derivative with respect to $z$ and as before where we have defined $\alpha = \log {\tilde a}$ and $\beta = \log c$.

We found that the asymptotic form of the dilaton at small $z$ had an important effect on the values of the entropy and mass of the black hole thus one needs to carefully study the effects of various dilaton boundary conditions to mimic the lattice QCD equation of state.
In the asymptotic regime near for small $z$  we seek  a black hole solution to to the equations of motion with a small asymptotic value for the dilaton field $\phi(z)$.

For small $z$ we set:
\[
\begin{array}{l}
 \tilde a = \frac{1}{z}e^{-\sigma \frac{1}{36}z^2} \\ 
 b = \frac{1}{z^2}e^{-\sigma \frac{1}{36}z^2}(1 -\mu z^2  + \frac{\tilde \lambda }{12 z^2})^{-1/2} \\ 
 c = e^{-\sigma \frac{1}{36}z^2}(1 -\mu z^2  + \frac{\tilde \lambda }{12 z^2})^{1/2} \\
 \end{array}
\]
where $\sigma$ is a deformation parameter. In terms of $\alpha$ and $\beta$ this becomes:
\[
\begin{array}{l}
 \alpha = - \log z -\sigma \frac{1}{36} z^2\\ 
 \beta = \frac{1}{2}\log(1-\mu z^2  + \frac{\tilde \lambda }{12 z^2})-\sigma \frac{1}{36}z^2\\
 \end{array}
\]
One can then use the second equation in (5.3) to solve for the asymptotic form of the dilaton by integrating:
\[
\phi' = \sqrt{- 6\alpha '' + 6\alpha '^2}
\]
 as in \cite{dePaula:2008fp}. This leads to the asymptotic form of the dilaton:
\[
\phi = \frac{1}{6\sqrt{6}}\sqrt{\sigma} z \sqrt{ 54 + z^2 \sigma} + 54 \frac{1}{6\sqrt{6}}\sinh^{-1}( \frac {z \sqrt{\sigma}}{3\sqrt{6}}).
\]
Expanding this expression for small $z$ we find:
\[
\phi = \sqrt{\sigma} z + \frac{1}{324} \sigma^{3/2} z^3 - \frac{1}{116640}\sigma^{5/2}z^5 + \frac{1}{17635968}\sigma^{7/2} z^7 + \dots
\]
One can use this expression to define a boundary condition on the dilaton field for small $z$. Then one can numerically solve for the black hole solution with potential (5.2) and a given deformation parameter $\sigma$. As $\sigma$ goes to zero one has the vacuum black hole solution discussed in the previous subsection.

\begin{figure}[tbp]
   \centerline{\hbox{
   \epsfxsize=4.0in
   \epsffile{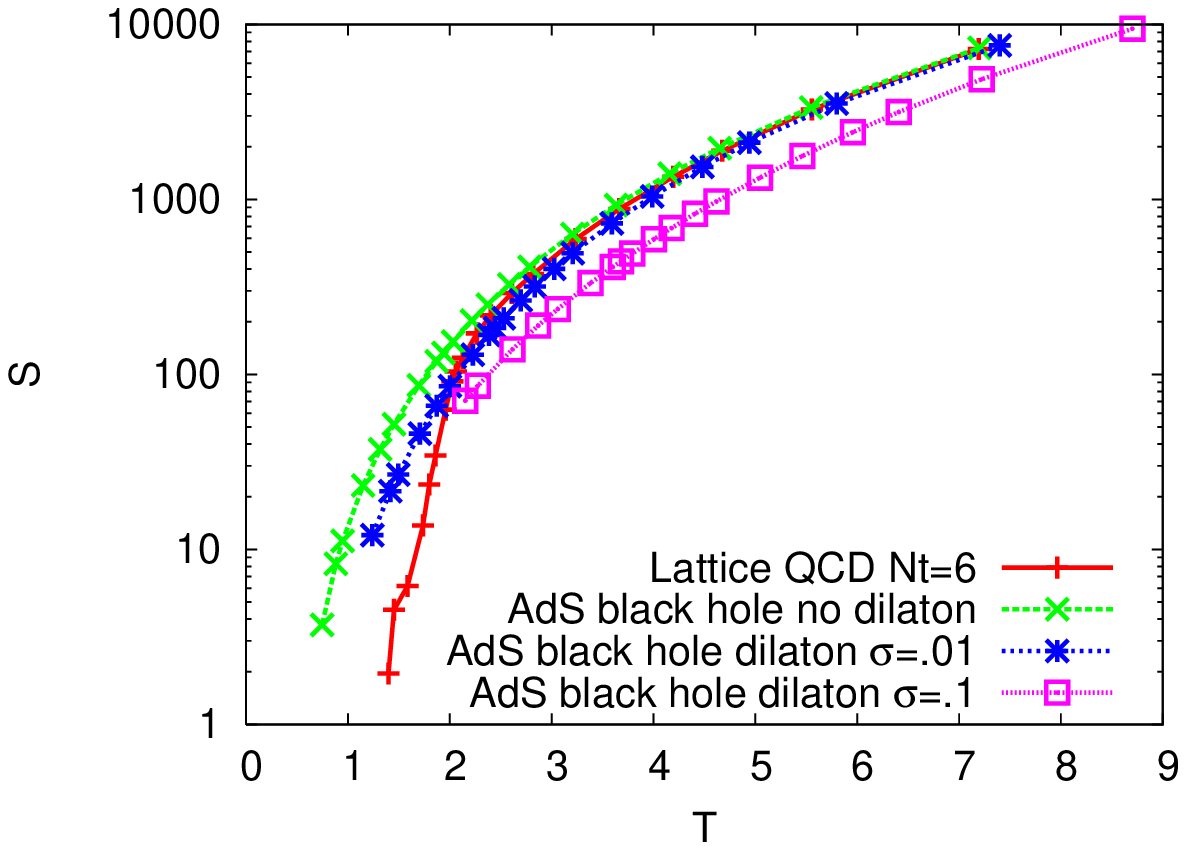}
     }}
   \centerline{\hbox{
   \epsfxsize=4.0in
   \epsffile{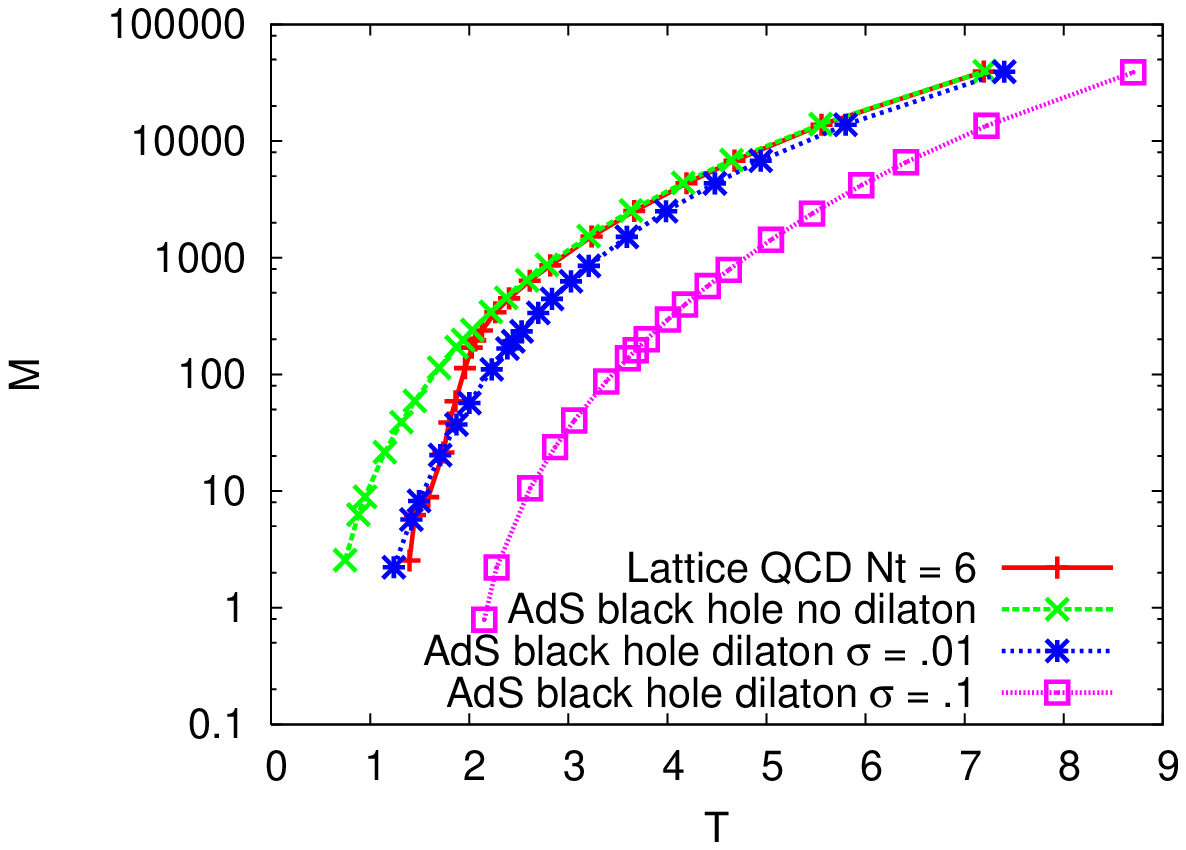}
     }}
   \caption{Comparison of entropy and mass (energy density) between
     Lattice QCD with $Nt = 6$ and $AdS$ black hole solutions with and
     without the dilaton.}
  \label{fig11}
\end{figure}

\begin{figure}[tbp]
   \centerline{\hbox{
   \epsfxsize=3.0in
   \epsffile{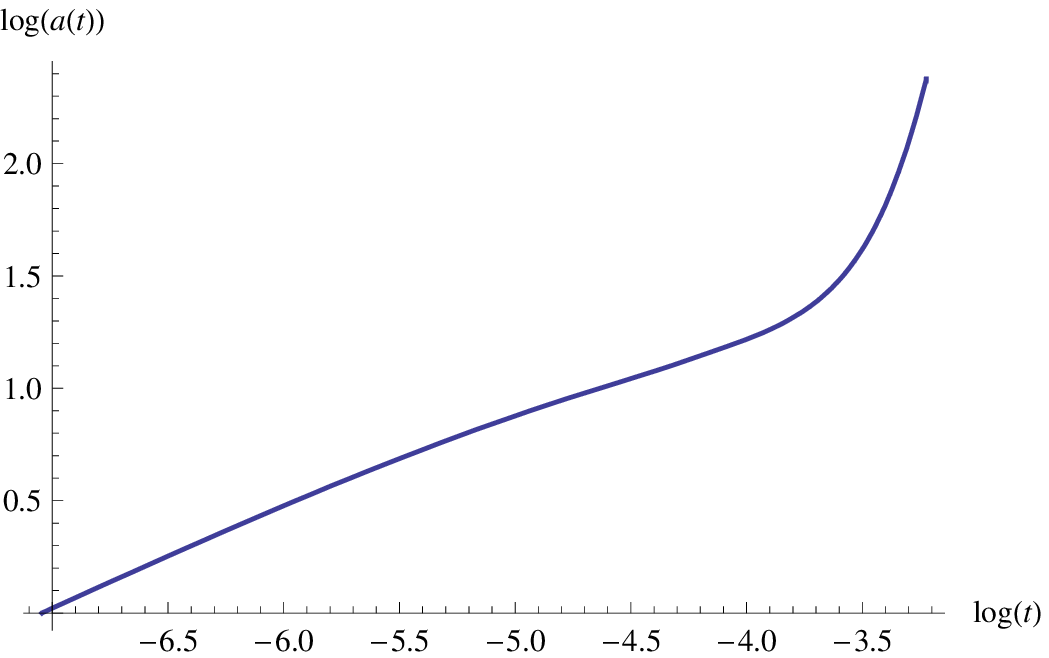}
   \epsfxsize=3.0in
   \epsffile{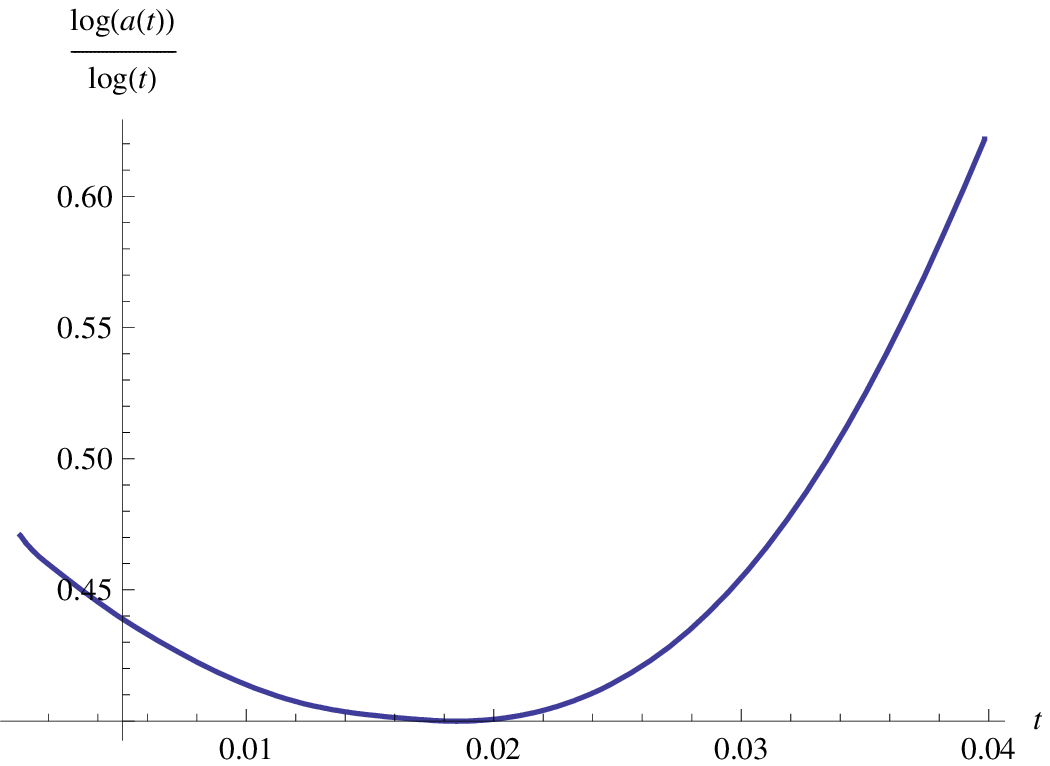}
     }
  }
  \caption{Scale factor as a function of log time from $AdS$ black hole equation of state with dilaton.}
  \label{fig12}
\end{figure}

For $\sigma = .01$ the entropy, mass (energy density), and temperature
are all modified by the dilaton. In Figure~\ref{fig11} we plot the
entropy and mass as a function of temperature for $\sigma = .01$. One
can then use equation~(\ref{eq:dTdts}) to calculate the time
dependence of the physical radius $a(t)$.  This is shown in
Figure~\ref{fig12}. The inclusion of the dilaton creates nontrivial
deviation from the radiation type expansion at late times and low
temperatures. In particular, we observe in Figure~\ref{fig12} the
upward bending hockey stick behavior that is qualitatively similar to
that of lattice QCD.

Without the dilaton the scale factor $a(t)$ can be described by the radiation expression $t^{1/2}$ over the temperature range covered by the lattice QCD calculation. With the dilaton it is possible to describe a late time behavior similar to lattice QCD. 

\section{Conclusion}

We have computed the time dependence of the scale factor from the
lattice QCD equation of state with $N_t=6$. We find that the scale
factor is described by a radiation dominated universe $t^{1/2}$ at
early times with a complicated time dependence at late times which
seems to be closer to a matter dominated universe. We compared our
results from lattice QCD with other approaches to the low temperature
equation of state including the hadronic resonance gas model, the
Hagedorn model and the AdS/CFT equation of state.

We found that on a log log plot the scale factor displayed an upward
pointing hockey stick behavior for the lattice QCD data. For the
hadronic resonance gas model over the temperature range of lattice QCD
data, the HRG model lies above the lattice QCD
data~\cite{Cheng:2007jq}. The slope of the log log plot for the scale
factor is between the radiation value .5 and the matter dominated
value .666 for the HRG model in this regime. The Hagedorn model leads
to a limiting temperature as a function of time and works better a low
temperature and late times. The equations are simpler than the HRG
model because less experimental input is required but lead to a
diverging energy density at the Hagedorn temperature unlike the
lattice QCD data. The Hagedorn cosmological model is still of interest
for theories where the limiting temperature is at the TeV scale or the
Planck scale. For the AdS/QCD model we found that without the dilaton
the entropy and energy density led to a $t^{1/2}$ expansion over the
temperature range of lattice QCD. We introduced a parameter $\sigma$
that determined the boundary condition at infinity for the dilaton.
Turning the dilaton on and tuning the parameter $\sigma$ allowed us to
derive a time dependence for the scale factor that was qualitatively
similar to that of lattice QCD through the AdS/QCD correspondence.

With future improvements in lattice QCD calculations and deeper understanding of the role of AdS/QCD it will be interesting to revisit the subject of QCD cosmology by adding the effects of bulk viscosity, chemical potentials and interaction with leptons. Eventually one would like to make contact with astrophysical measurements of  the early Universe as envisioned in early references such as \cite{Witten:1984rs}.

\section*{Acknowledgments}
We wish to thank Frithjof Karsch, Michael Creutz, Dmitri Kharzeev and Mikko Laine for useful discussions and suggestions.
This manuscript has been authored in part by Brookhaven Science Associates, LLC, under Contract No. DE-AC02-98CH10886 with the U.S. Department of Energy.


\begin{thebibliography}{100}

\bibitem{Witten:1984rs}
  E.~Witten,
  ``Cosmic Separation Of Phases,''
  Phys.\ Rev.\  D {\bf 30}, 272 (1984).

\bibitem{Schwarz:2003du}
  D.~J.~Schwarz,
  ``The first second of the universe,''
  Annalen Phys.\  {\bf 12}, 220 (2003)
  [arXiv:astro-ph/0303574].

\bibitem{Muller}
  L.~M.~M\"uller, Master's thesis, ``The quark-hadron
  phase transition in the early Universe'', (2001).

\bibitem{Borghini:2000yp}
  N.~Borghini, W.~N.~Cottingham and R.~Vinh Mau,
  ``Possible cosmological implications of the quark-hadron phase  transition,''
  J.\ Phys.\ G {\bf 26}, 771 (2000)
  [arXiv:hep-ph/0001284].

\bibitem{Kapusta:2000fe}
  J.~I.~Kapusta,
  ``Quark-gluon plasma in the early universe,''
  arXiv:astro-ph/0101516.

\bibitem{Chandra:1999tr}
  D.~Chandra and A.~Goyal,
  ``Dynamical evolution of the universe in the quark-hadron phase  transition
  and possible nugget formation,''
  Phys.\ Rev.\  D {\bf 62}, 063505 (2000)
  [arXiv:hep-ph/9903466].

\bibitem{Coley:1993zg}
  A.~A.~Coley and T.~Trappenberg,
  ``The Quark - hadron phase transition, QCD lattice calculations and
  inhomogeneous big bang nucleosynthesis,''
  Phys.\ Rev.\  D {\bf 50}, 4881 (1994)
  [arXiv:astro-ph/9307031].

\bibitem{Suganuma:1996yb}
  H.~Suganuma, H.~Ichie, H.~Monden, S.~Sasaki, M.~Orito, T.~Yamamoto and T.~Kajino,
  ``QCD phase transition at high temperature in cosmology,''
  arXiv:hep-ph/9608333.

\bibitem{Olive:1990rd}
  K.~A.~Olive,
  ``The Quark - hadron transition in cosmology and astrophysics,''
  Science {\bf 251}, 1194 (1991).

\bibitem{Hagedorn:1965st}
  R.~Hagedorn,
  ``Statistical thermodynamics of strong interactions at high-energies,''
  Nuovo Cim.\ Suppl.\  {\bf 3}, 147 (1965).

\bibitem{Huang:1970iq}
  K.~Huang and S.~Weinberg,
  ``Ultimate temperature and the early universe,''
  Phys.\ Rev.\ Lett.\  {\bf 25}, 895 (1970).

\bibitem{Cheng:2007jq}
  M.~Cheng {\it et al.},
  ``The QCD Equation of State with almost Physical Quark Masses,''
  Phys.\ Rev.\  D {\bf 77}, 014511 (2008)
  [arXiv:0710.0354 [hep-lat]].

\bibitem{Gubser:2008ny}
  S.~S.~Gubser and A.~Nellore,
  ``Mimicking the QCD equation of state with a dual black hole,''
  arXiv:0804.0434 [hep-th].

\bibitem{Gubser:2008yx}
  S.~S.~Gubser, A.~Nellore, S.~S.~Pufu and F.~D.~Rocha,
  ``Thermodynamics and bulk viscosity of approximate black hole duals to finite
  temperature quantum chromodynamics,''
  arXiv:0804.1950 [hep-th].

\bibitem{Gubser:2008sz}
  S.~S.~Gubser, S.~S.~Pufu and F.~D.~Rocha,
  ``Bulk viscosity of strongly coupled plasmas with holographic duals,''
  arXiv:0806.0407 [hep-th].

\bibitem{Freedman:1977gz}
  B.~Freedman and L.~D.~McLerran,
  ``Quark Star Phenomenology,''
  Phys.\ Rev.\  D {\bf 17}, 1109 (1978).

\bibitem{Chakrabarty:1991ui}
  S.~Chakrabarty,
  ``Equation of state of strange quark matter and strange star,''
  Phys.\ Rev.\  D {\bf 43}, 627 (1991).

\bibitem{Yao:2006px}
  W.~M.~Yao {\it et al.}  [Particle Data Group],
  ``Review of particle physics,''
  J.\ Phys.\ G {\bf 33}, 1 (2006).

\bibitem{Luzum:2008cw}
  M.~Luzum and P.~Romatschke,
  ``Conformal Relativistic Viscous Hydrodynamics: Applications to RHIC,''
  arXiv:0804.4015 [nucl-th].

\bibitem{Kharzeev:2007wb}
  D.~Kharzeev and K.~Tuchin,
  ``Bulk viscosity of QCD matter near the critical temperature,''
  arXiv:0705.4280 [hep-ph].

\bibitem{Karsch:2003vd}
  F.~Karsch, K.~Redlich and A.~Tawfik,
  ``Hadron resonance mass spectrum and lattice QCD thermodynamics,''
  Eur.\ Phys.\ J.\  C {\bf 29}, 549 (2003)
  [arXiv:hep-ph/0303108].

\bibitem{Karsch:2003zq}
  F.~Karsch, K.~Redlich and A.~Tawfik,
  ``Thermodynamics at non-zero baryon number density: A comparison of  lattice
  and hadron resonance gas model calculations,''
  Phys.\ Lett.\  B {\bf 571}, 67 (2003)
  [arXiv:hep-ph/0306208].


\bibitem{SakthiMurugesan:1990pe}
  K.~Sakthi Murugesan, G.~Janhavi and P.~R.~Subramanian,
  ``Can the phase transition from quark - gluon plasma to hadron resonance gas
  affect primordial nucleosynthesis?,''
  Phys.\ Rev.\  D {\bf 41}, 2384 (1990).

\bibitem{Tawfik:2004sw}
  A.~Tawfik,
  ``The QCD phase diagram: A comparison of lattice and hadron resonance gas
  model calculations,''
  Phys.\ Rev.\  D {\bf 71}, 054502 (2005)
  [arXiv:hep-ph/0412336].

\bibitem{Shifman:2008ja}
  M.~Shifman and M.~Unsal,
  ``QCD-like Theories on $R_3\times S_1$: a Smooth Journey from Small to Large
  $r(S_1)$ with Double-Trace Deformations,''
  arXiv:0802.1232 [hep-th].

\bibitem{Gupta:2008} R.~Gupta, ``The EOS from simulations on BlueGene
  L Supercomputer at LLNL and NYBlue,``
  PoS {\bf LAT2008} 170 (2008)


\bibitem{Laine:2006cp}
  M.~Laine and Y.~Schroder,
  ``Quark mass thresholds in QCD thermodynamics,''
  Phys.\ Rev.\  D {\bf 73}, 085009 (2006)
  [arXiv:hep-ph/0603048].

\bibitem{Cheng:2007wu}
  M.~Cheng  [RBC-Bielefeld Collaboration],
  ``Charm Quarks and the QCD Equation of State,''
  PoS {\bf LAT2007}, 173 (2007)
  [arXiv:0710.4357 [hep-lat]].

\bibitem{Endrodi:2007tq}
  G.~Endrodi, Z.~Fodor, S.~D.~Katz and K.~K.~Szabo,
  ``The equation of state at high temperatures from lattice QCD,''
  PoS {\bf LAT2007}, 228 (2007)
  [arXiv:0710.4197 [hep-lat]].

\bibitem{Miller:2006hr}
  D.~E.~Miller,
  ``Lattice QCD calculation for the physical equation of state,''
  Phys.\ Rept.\  {\bf 443}, 55 (2007)
  [arXiv:hep-ph/0608234].


\bibitem{BraunMunzinger:2003zd}
  P.~Braun-Munzinger, K.~Redlich and J.~Stachel,
  ``Particle production in heavy ion collisions,''
  arXiv:nucl-th/0304013;
  A.~Andronic, P.~Braun-Munzinger and J.~Stachel,
  ``Hadron production in central nucleus nucleus collisions at chemical
  freeze-out,''
  Nucl.\ Phys.\  A {\bf 772}, 167 (2006)
  [arXiv:nucl-th/0511071].

\bibitem{Carlitz:1972uf}
  R.~D.~Carlitz,
  ``Hadronic matter at high density,''
  Phys.\ Rev.\  D {\bf 5}, 3231 (1972).

\bibitem{Castorina:2007eb}
  P.~Castorina, D.~Kharzeev and H.~Satz,
  ``Thermal Hadronization and Hawking-Unruh Radiation in QCD,''
  Eur.\ Phys.\ J.\  C {\bf 52}, 187 (2007)
  [arXiv:0704.1426 [hep-ph]].


\bibitem{Harmark:2006ta}
  T.~Harmark and M.~Orselli,
  ``Matching the Hagedorn temperature in AdS/CFT,''
  Phys.\ Rev.\  D {\bf 74}, 126009 (2006)
  [arXiv:hep-th/0608115].


\bibitem{KalyanaRama:1998cb}
  S.~Kalyana Rama and B.~Sathiapalan,
  ``The Hagedorn transition, deconfinement and the AdS/CFT correspondence,''
  Mod.\ Phys.\ Lett.\  A {\bf 13}, 3137 (1998)
  [arXiv:hep-th/9810069].


\bibitem{Antoniadis:1999fj}
  I.~Antoniadis and B.~Pioline,
  ``Large dimensions and string physics at a TeV,''
  arXiv:hep-ph/9906480.


\bibitem{Nayeri:2005ck}
  A.~Nayeri, R.~H.~Brandenberger and C.~Vafa,
  ``Producing a scale-invariant spectrum of perturbations in a Hagedorn  phase
  of string cosmology,''
  Phys.\ Rev.\ Lett.\  {\bf 97}, 021302 (2006)
  [arXiv:hep-th/0511140].

\bibitem{Tyson}
P. Tyson, ``Absolute Hot: Is there an opposite to absolute zero?''\\
http://www.pbs.org/wgbh/nova/zero/hot.html


\bibitem{Poisson:1990eh}
  E.~Poisson and W.~Israel,
  ``Internal structure of black holes,''
  Phys.\ Rev.\  D {\bf 41}, 1796 (1990).

\bibitem{Fischler:1990pk}
  W.~Fischler, D.~Morgan and J.~Polchinski,
  ``Quantization of false vacuum bubbles:A Hamiltonian treatment of gravitational tunneling,''
  Phys.\ Rev.\  D {\bf 42}, 4042 (1990).

\bibitem{Pidokrajt}
N. Pidokrajt, ``Black hole thermodynamics'', master's thesis (2003).

\bibitem{Louko:1996jd}
  J.~Louko, J.~Z.~Simon and S.~N.~Winters-Hilt,
  ``Hamiltonian thermodynamics of a Lovelock black hole,''
  Phys.\ Rev.\  D {\bf 55}, 3525 (1997)
  [arXiv:gr-qc/9610071].

\bibitem{Quevedo:2008xn}
  H.~Quevedo and A.~Sanchez,
  ``Geometrothermodynamics of asymptotically de Sitter black holes,''
  arXiv:0805.3003 [hep-th].

\bibitem{dePaula:2008fp}
  W.~de Paula, T.~Frederico, H.~Forkel and M.~Beyer,
  ``Dynamical AdS/QCD with area-law confinement and linear Regge
  trajectories,''
  arXiv:0806.3830 [hep-ph].


\end{thebibliography}
\end{document}